\documentclass[times,twocolumn,final,nopreprintline]{elsarticle}

\usepackage{framed,multirow}

\usepackage{amssymb}
\usepackage{latexsym}
\usepackage{amsmath}
\usepackage{graphicx}
\usepackage{subcaption}
\usepackage{float}
\usepackage{tabularx}

\usepackage{url}
\usepackage{xcolor}

\usepackage{hyperref}
\usepackage[capitalise]{cleveref}

\definecolor{newcolor}{rgb}{.8,.349,.1}

\def\A{\mathbf A}
\def\F{\mathcal F}
\def\x{\mathbf x}
\def\y{\mathbf y}
\def\z{\mathbf z}
\def\h{\mathbf h}
\def\k{\mathbf k}
\def\M{\mathbf M}
\def\v{\mathbf v}
\def\e{\mathbf e}
\def\V{\mathbf V}
\def\Lambdab{\boldsymbol \Lambda}
\def\chib{\boldsymbol \chi}

\begin{document}

\begin{frontmatter}

\title{DeepSTI: Towards Tensor Reconstruction using Fewer Orientations  in Susceptibility Tensor Imaging}

\author[1]{Zhenghan {\color{black}Fang}}
\author[1]{Kuo-Wei {\color{black}Lai}}
\author[2,3]{Peter {\color{black}van Zijl}}
\author[2,3]{{Xu {\color{black}Li}}\corref{cor1}}
\author[1]{Jeremias {\color{black}Sulam}\corref{cor1}}
\cortext[cor1]{Corresponding authors: Jeremias Sulam, email: jsulam1@jhu.edu; Xu Li, email: xuli@mri.jhu.edu.}

\address[1]{
Department of Biomedical Engineering, Johns Hopkins University, Baltimore, MD 21218, USA}
\address[2]{F.M. Kirby Research Center for Functional Brain Imaging, Kennedy Krieger Institute, Baltimore,  MD 21205, USA}
\address[3]{Department of Radiology and Radiological Sciences, Johns Hopkins University, Baltimore, MD  21205, USA}

\begin{abstract}
Susceptibility tensor imaging (STI) is an emerging magnetic resonance imaging technique that characterizes the anisotropic tissue magnetic susceptibility with a second-order tensor model. STI has the potential to provide information for both the reconstruction of white matter fiber pathways and detection of myelin changes in the brain at mm resolution or less, which would be of great value for understanding brain structure and function in healthy and diseased brain. However, the application of STI in vivo has been hindered by its cumbersome and time-consuming acquisition requirement of measuring susceptibility induced MR phase changes at multiple head orientations. Usually, sampling at more than six orientations is required to obtain sufficient information for the ill-posed STI dipole inversion. This complexity is enhanced by the limitation in head rotation angles due to physical constraints of the head coil. As a result, STI has not yet been widely applied in human studies in vivo. In this work, we tackle these issues by proposing an image reconstruction algorithm for STI that leverages data-driven priors. Our method, called DeepSTI, learns the data prior implicitly via a deep neural network that approximates the proximal operator of a regularizer function for STI. The dipole inversion problem is then solved iteratively using the learned proximal network. Experimental results using both simulation and in vivo human data demonstrate great improvement over state-of-the-art  algorithms in terms of the reconstructed tensor image, principal eigenvector maps and tractography results, while allowing for tensor reconstruction with MR phase measured at much less than six different orientations. Notably, promising reconstruction results are achieved by our method from only one orientation in human in vivo, and we demonstrate a potential application of this technique for estimating lesion susceptibility anisotropy in patients with multiple sclerosis.

\end{abstract}

\begin{keyword}
Susceptibility tensor imaging\sep
Proximal learning\sep
Deep learning\sep
Dipole inversion\sep
In vivo human brain\sep
Fiber pathways\sep
Fiber tractography\sep 
Myelin imaging
\end{keyword}

\end{frontmatter}


\section{Introduction}
\label{sec1}
The characterization of neural fiber pathways in the brain is important for understanding brain function and development, as well as for diagnosis and treatment of various neurological and psychiatric diseases \cite{glasser2016human}. Estimation of local white matter fiber directions and subsequent tracking of fiber pathways allows for noninvasive characterization of whole-brain neural networks as well as physical connections between different brain regions. Furthermore, it is often of great interest to understand not only the anatomical aspects of white matter fibers, but also their pathophysiological condition in terms of axonal integrity, myelination, and tissue composition, as those conditions have important implications for the study of brain development \citep{gilles1983myelinated,van1991myelination,pujol2006myelination,nave2014myelination,monje2018myelin,grotheer2022white}, and a variety of neurodegenerative diseases, for instance multiple sclerosis (MS) \citep{langkammer2013quantitative,chen2014quantitative,li2016magnetic,wiggermann2017susceptibility} and Alzheimer's disease \citep{acosta2013vivo,bouhrara2018evidence,ayton2017cerebral,kim2017quantitative,chen2021quantitative}. 

The development of diffusion tensor imaging (DTI), has allowed for noninvasive mapping of 3D fiber pathways of  the human brain in vivo \citep{mori2002fiber,wakana2004fiber,jung20093d,schilling2018histological,jeurissen2019diffusion,sotiropoulos2019building}. Broadly speaking, DTI, and more advanced approaches such as multi-shell multi-orientation diffusion mapping, track white matter fibers by imaging the restricted movement of water molecules inside nerve fibers \citep{jeurissen2014multi,jeurissen2019diffusion}. DTI still is the most commonly used method for imaging neural fiber pathways for clinical applications.
While recent advances in pulse sequence design and scanner gradients \cite{mcnab2013human,fan2014investigating} allow resolutions of about 1.5mm isotropic \cite{glasser2016human}, most clinical diffusion MRI studies still are acquired with relative poor spatial resolution, with typical imaging voxel sizes around 2-2.5 mm isotropic. In addition, diffusion MRI is mainly sensitive to the axon cell membrane barrier with much less sensitivity to myelination changes in the brain \citep{beaulieu2002basis,liu2011high}, which is of specific clinical interest.

A promising alternative to DTI-based tractography is susceptibility tensor imaging (STI) \cite{liu2010susceptibility}. Unlike DTI, STI measures the anisotropic magnetic susceptibility of brain tissue originating from aligned tissue micro-structures with anisotropic molecular susceptibility, e.g. the white matter myelin sheath \cite{li2012magnetic,li2017susceptibility}. Using a gradient echo (GRE) sequence to measure the susceptibility induced MR phase changes, STI has the potential to achieve much higher spatial resolution (sub-millimeter) in imaging the fiber pathways than DTI, especially at high magnetic field strengths (3T and above). In addition, because tissue magnetic susceptibility is sensitive to changes in tissue components such as myelin and iron \citep{sibgatulin2021vivo}, STI derived measures can potentially provide new insights into the pathophysiological processes during brain development and degeneration as shown in fetal alcohol spectrum disorder \citep{cao2014prenatal} and multiple sclerosis \citep{wisnieff2015quantitative,sibgatulin2022magnetic,li2016magnetic,wisnieff2013magnetic,haacke2015quantitative,liu2015susceptibility}.

Yet, current STI techniques suffer from long scanning time and impractical protocols, which pose major barriers for routine application of human STI in vivo. Obtaining one STI scan requires rotating the subject's head multiple times inside the magnet and acquiring a series of measurements at each head orientation. Assuming the susceptibility tensor is real and symmetric, at least six orientations need to be acquired to solve the ill-posed dipole inversion problem for reconstruction of a STI image \citep{liu2010susceptibility}---and several more are often desirable to improve image quality.  Holding the head at non-supine positions for a long time is difficult and increases discomfort for patients. Beyond this, constraints based on head coil size limit the possible range of head rotation angles to typically within 25 degrees from the direction of main magnetic field for \textit{in vivo} human experiments \citep{li2012mapping}. Such a limitation results in a high condition number for the STI forward model, posing significant challenges for reliable image reconstruction even when a large number of orientations can be acquired\citep{li2014mean, bao2021diffusion}. As a result, STI is yet to achieve fiber tracking results as consistent and reliable as DTI, especially for in vivo human measurements \cite{bilgic2016rapid}.

\begin{figure}
    \centering
    \includegraphics[trim = 5 22 10 5, width=.47\textwidth]{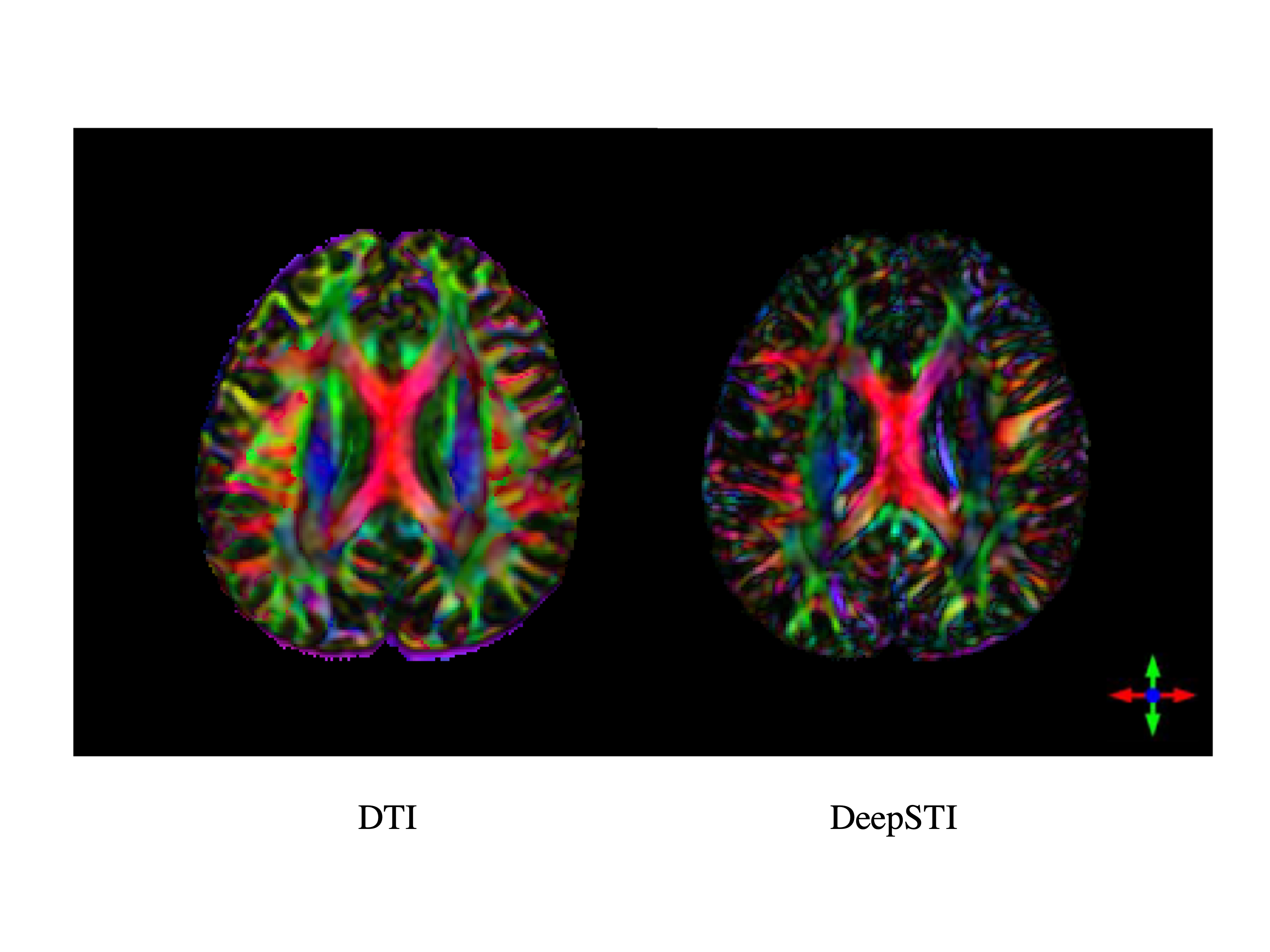}\\[12pt]
    \caption{Anisotropy weighted principal eigenvector (PEV) maps estimated from real measurements from a human subject. Left: fractional anisotropy (FA) weighted DTI PEV map acquired at 3T and 2.2 mm isotropic resolution (interpolated to 0.98$\times$0.98$\times$1 mm resolution). Right: magnetic susceptibility anisotropy (MSA) weighted STI PEV map estimated by DeepSTI from 4 GRE measurements at 7T acquired at 1 mm isotropic resolution (reconstructed to 0.98$\times$0.98$\times$1 mm resolution). The acquisition time for STI was 21 minutes (4 head orientations, 5 minutes 15 seconds per orientation). The acquisition time of DTI was 8 minutes 50 seconds.}
    \label{fig:res-pev-real-7T}
\end{figure}

In this paper, we present an approach towards solving the above issues by developing new machine learning based methods for STI image reconstruction that allow for a reduced number of head orientations. Specifically, we developed a STI algorithm, called DeepSTI, that approximately minimizes a regularized STI reconstruction problem, where the regularization is based on a prior that is learned from data. To this end, we leveraged advances in deep learning methods, which have recently demonstrated encouraging results in solving ill-posed inverse problems in medical imaging but have not yet been applied for the challenging problem of STI reconstruction, potentially due to its high-dimensional nature and the difficulty to incorporate MR phase measures acquired at varying number of orientations and arbitrary angles as input. 
Here, we exploited the learned proximal convolutional neural network (LPCNN) model, used previously for reconstructing \emph{isotropic} magnetic susceptibility sources as in quantitative susceptibility mapping (QSM) \citep{lai2020learned}, and we expanded it to bear in the context of this more challenging task of STI. We validated our method using simulations as well as applied it on in vivo human data, and compared it to existing state-of-the-art STI reconstruction methods. The proposed algorithm shows unprecedented results for STI reconstruction with fewer than 6 orientations, with superior performance over previous methods in the reconstruction of tensor image and principal eigenvector (PEV) maps. DeepSTI also produces more coherent results in fiber tractography from STI, and is not limited to using uniform resolution across samples. \Cref{fig:res-pev-real-7T} depicts an example result, demonstrating the improved resolution of STI over that of DTI, on a human subject. 

The rest of paper is organized as follows. Section \ref{sec:related-works} summarizes related works. Section \ref{sec:methods} describes our proposed method for STI reconstruction, while Section \ref{sec:results} presents experiments and results. Section \ref{sec:discussion} includes a further discussion on advantages and limitations of our method before concluding in Section \ref{sec:conclusion}

\section{Related work}
\label{sec:related-works}
A variety of methods have been proposed for STI reconstruction \citep{liu2010susceptibility,liu20123d,li2014mean,bilgic2016rapid,dibb2017joint,bao2021diffusion,cao2021asymmetric}. The original approach \citep{liu2010susceptibility} used a least square solver to tackle the dipole inversion problem in the frequency domain. Subsequently, \cite{liu20123d} proposed to penalize high spatial frequency anisotropic components, leading to improvements in reconstruction and less sensitivity to imperfect image registration. However, this can lead to over-smoothing \citep{cao2021asymmetric}, undermining the advantages of STI for high-resolution fiber tracking. Another study \citep{li2014mean} proposed to limit the magnetic susceptibility anisotropy (MSA) to white matter regions as determined from a coregistered T1-weighted image and imposed morphology constraints on the mean magnetic susceptibility (MMS), allowing more accurate tensor estimation. Yet, this approach is naturally sensitive to errors in white matter segmentation and imposes a strong assumption on the anisotropy distribution. Joint estimation of mutually anisotropic relaxation tensor and susceptibility tensor regions from the same multi-echo acquisitions has also been proposed to better determine the principal eigenvectors of the susceptibility tensors \citep{dibb2017joint}. More recently, \cite{bao2021diffusion} proposed diffusion-regularized STI using the fiber directions estimated by DTI to regularize the STI solution. When using only 6 orientations, this approach achieved better reconstruction than traditional methods and displayed reduced artifacts in fiber direction maps. However, this method requires additional acquisition time for DTI and the regularization from low-resolution DTI can lead to over-smoothed STI solutions. Finally, \cite{cao2021asymmetric} proposed the asymmetric susceptibility tensor imaging (aSTI) approach in which the susceptibility tensor is fitted with 9 unknowns first and then decomposed into the symmetric and anti-symmetric components. This approach resulted in less noise and streaking artifacts in the reconstructed symmetric susceptibility tensor, while the anti-symmetric components captured mainly noise and artifacts. 
Further MMS and magnetic susceptibility anosotropy (MSA) based regularizations of the aSTI approach have also been tested very recently \citep{shi2022regularized}.
Overall, all these approaches require MR phase images acquired at a large number of head orientations for satisfactory reconstruction, which hampers the wide application of STI in human studies in vivo.
On the other hand, a number of machine learning based approaches have been proposed for dipole inversion in the simpler problem of QSM \citep{yoon2018quantitative,bollmann2019deepqsm,kames2019proximal,jung2020exploring,chen2020qsmgan,polak2020nonlinear,lai2020learned,jung2022overview}, where the susceptibility is assumed to be isotropic and only the equivalent MMS is estimated. These methods are encouraging, but have not yet been extended to the more challenging setting of STI estimation. 

In this work we leverage tools from learned proximal neural networks. The idea of learning a proximal operator using deep learning has been explored previously in a few studies for inverse tasks in medical imaging. For example, \cite{adler2018learned} proposed a Learned Primal-Dual algorithm for tomographic reconstruction, which used a modified version of the primal dual hybrid gradient algorithm \citep{chambolle2011first} replacing the proximal operator by convolutional neural networks learned from training data.  \cite{mardani2018deep} proposed to use generative adversarial networks (GAN) for compressed sensing MRI, leveraging the generator for aliasing removal by projecting an initial aliased estimate onto the learned low-dimensional manifold of high-quality MR images. Finally, in the related work of \cite{lai2020learned} the authors proposed a learned proximal CNN method for QSM reconstruction (LPCNN-QSM), training a Wide ResNet \citep{zagoruyko2016wide} as the proximal operator within the proximal gradient descent algorithm for solving the dipole inversion. LPCNN-QSM \citet{lai2020learned} not only achieved state-of-the-art reconstruction accuracy, but also naturally handled an arbitrary number of phase inputs at arbitrary orientations with a single trained model, providing far more flexibility than other methods.

However, difficulties remain in applying learned proximal models in STI. First, the condition number of the forward operator in STI is high ($> 50$, \citep{bao2021diffusion}), which makes its inverse problem more ill-posed than QSM and the solution less stable and more sensitive to noise. Second, unlike QSM, where high-quality reconstructions can be obtained from human subjects in vivo through multiple orientation sampling (COSMOS) \citep{liu2009calculation}, high-quality ground-truth images for human STI are almost impossible to obtain with current technology, posing challenges for both training and evaluation. Furthermore, the underlying image dimension of the STI problem is 6 times higher than QSM. This allows quantification of susceptibility anisotropy and white matter fiber orientations, but it also increases the number of unknowns and makes training more difficult, both statistically and computationally (including GPU memory constraints). Finally, the need for incorporating a varying number of measurements at arbitrary orientation angles as input during training (and testing) is an additional difficulty for data-driven STI reconstruction methods in general.

\section{Methods}
\label{sec:methods}
When a material is placed in a magnetic field, it gets magnetized and creates a local magnetic field perturbation. The ratio of its magnetization to the applied magnetic field is defined as the \emph{magnetic susceptibility} \citep{liu2010susceptibility,li2017susceptibility}. For a specific anisotropic material, e.g. white matter fibers, the magnetic susceptibility can be approximated by a real symmetric 3$\times$3 tensor \citep{li2014mean,li2017susceptibility,bao2021diffusion}, denoted by $\chib$:
\begin{align}
    \chib = 
    \begin{bmatrix}
    \chi_{11} & \chi_{12} & \chi_{13} \\
    \chi_{12} & \chi_{22} & \chi_{23} \\
    \chi_{13} & \chi_{23} & \chi_{33}
    \end{bmatrix} \in \mathbb{R}^{3\times 3}.
\end{align}
For axial symmetry, e.g. in cylindrial structures, this tensor contains 6 independent elements: $\chi_{11}$, $\chi_{12}$, $\chi_{13}$, $\chi_{22}$, $\chi_{23}$ and $\chi_{33}$.

The goal of STI is to characterize the magnetic susceptibility tensor of tissue at every voxel. Assuming each voxel can be described by a 3$\times$3 symmetric tensor, each susceptibility tensor image can be divided into 6 sub-volumes, where each sub-volume corresponds to one independent tensor element. Denote the $\chi_{ij}$ element at the $v$-th voxel as $\chi^v_{ij}$. Then, each sub-volume is given by
\begin{align}
    \x_{ij} = \begin{bmatrix}\chi^1_{ij} \dots \chi^v_{ij} \dots \chi^n_{ij} \end{bmatrix} \in \mathbb{R}^n,
\end{align}
where $i,j\in\{1,2,3\}$, $i \leq j$, and $n$ is the number of voxels.
Thus, the entire susceptibility tensor image $\x$ can be given by
\begin{align}
    \x = \begin{bmatrix} \x_{11}\ \x_{12}\ \x_{13}\ \x_{22}\ \x_{23}\ \x_{33}\end{bmatrix}^T \in \mathbb{R}^{6n}.
\end{align}

Magnetized materials generate their own magnetic field that in turn perturbs the main magnetic field. Such local field changes cause variations in the material's resonance frequency, which can be measured via phase images in MRI. The overall magnetic field perturbation, as measured by the phase image, represents the sum of all the field changes generated by each constituent magnetic dipole \citep{li2017susceptibility,ruetten2019introduction}, and can be calculated as the convolution of a susceptibility distribution with the magnetic field induced by a unit magnetic dipole \citep{liu2010susceptibility,liu2015susceptibility,li2017susceptibility,ruetten2019introduction}. 
Such convolution can be written in a simpler form in the spatial frequency domain using a Fourier Transform \citep{salomir2003fast,marques2005application}. Therefore, we can express the relationship between the normalized local field change, $\delta B \in \mathbb{R}^n$, and the underlying susceptibility tensor image $\x$ as:
\begin{align}\label{eq:local_phase_change}
    \delta B = \F^{-1}\A\F~\x + \e,
\end{align}
where $\F$ denotes the discrete Fourier Transform\footnote{
Note that the right-most Fourier operator in \cref{eq:local_phase_change} is applied to each tensor component $\x_{11},\x_{12},\dots,\x_{33}$ separately.}, $\e$ represents noise in the measurements, and $\A\in\mathbb{R}^{n\times 6n}$ is the dipole kernel in the frequency domain. The dipole kernel is composed of 6 sub-matrices:
\begin{align}
    \A = \begin{bmatrix}
    \A_{11}, ~ \A_{12}+\A_{21}, ~ \A_{13}+\A_{31}, ~ \A_{22}, ~ \A_{23}+\A_{32},~ \A_{33}
    \end{bmatrix},
\end{align}
where each sub-matrix $\A_{ij} \in \mathbb{R}^{n\times n}$ is diagonal and multiplies the Fourier transform of $\x_{ij}$. It can be shown \citep{liu2010susceptibility} that the diagonal elements of $A_{ij}$ are given by
\begin{equation}
A_{ij}(\k) = \frac{h_i h_j}{3} - \frac{\k^T\h  k_i h_j}{\k^T\k},
\end{equation}
where $\k = [k_1, k_2, k_3]^T \in \mathbb{R}^3$ is the spatial frequency vector and $\h = [h_1, h_2, h_3]^T \in \mathbb{R}^3$ is the direction of the main magnetic field in the subject frame of reference.

The above \cref{eq:local_phase_change} represents the field perturbation induced by the susceptibility tensor acquired at a specific brain orientation. 
For measurements from multiple head orientations, we then have
\begin{align}
    \begin{bmatrix}
    \delta B^1 \\
    \delta B^2 \\
    \vdots \\
    \delta B^m
    \end{bmatrix} = \begin{bmatrix}\F^{-1}\A^1 \\ \F^{-1}\A^2 \\ \vdots \\ \F^{-1}\A^m\end{bmatrix}\F~\x + \e = \mathcal F^{-1} \mathcal A \F \x + \e
\label{eq:forward}
\end{align}
where $\delta B^c$ and $\A^c$ are the normalized local field change and dipole kernel at the $c$-th orientation, respectively, and $m$ is the number of orientations. For simplicity, we denote $\mathcal A$ as the concatenation of the different dipole operators $\A^c$.

Given this forward model that characterizes the mapping from the susceptibility tensor image, $\x$, to the normalized local field change measurements ($\delta B^1, \dots , \delta B^m$), the goal of STI reconstruction, or dipole inversion, is to recover the former from the latter. Since the head rotation of human subjects in an MRI scanner is limited by the narrow head coil (to about $25^{\circ}$ away from the $z$-axis, the direction of main magnetic field), the condition number of the system matrix $\mathcal{A}$ in STI is large \citep{bao2021diffusion}, making the inverse solution exceptionally unstable, vulnerable to small errors and noise in the measurements. Hence, a good prior is critical for accurate estimation of $\x$. For simplicity, we henceforth denote the measurements at $m$ different orientations by \[\y = [\delta B^1,~\dots~, \delta B^m]^T \in \mathbb R^{nm}. \] 
Then, the dipole inversion problem can be written as the following regularized linear inverse problem:
\begin{align}
    \min_{\x} \frac{1}{2} \|\M(\y - \F^{-1}\mathcal{A}\F\x)\|_2^2 + R(\x),
    \label{eq:min}
\end{align}
where $\M$ a brain mask\footnote{$\M$ is a binary mask with $1$ inside the brain area and $0$ elsewhere.}, $f(\x) = \frac{1}{2} \|\M(\y- \F^{-1}\mathcal{A}\F\x)\|_2^2 : \mathbb{R}^{6n} \rightarrow \mathbb{R}$ is a data fidelity term that promotes the solution to match observed measurements $\y$, and $R(\x): \mathbb{R}^{6n} \rightarrow \mathbb{R}$ is a suitable regularization term that introduces prior knowledge\footnote{From a Bayesian perspective, a regularizer can be often interpreted as the negative logarithm of a prior distribution.} and allows a robust reconstruction for the ill-posed inverse problem \citep{benning2018modern}. All STI reconstruction methods \citep{liu20123d,li2014mean,bao2021diffusion} employ some form of the problem formulation in \cref{eq:min}, while differing in the choice of the prior, $R(\x)$. In what follows, we expand on the choice of prior used in this work, which will be learned from data.

\subsection{Learned Proximal Networks for STI Reconstruction}

In this work, the prior $R(\x)$ will be defined implicitly, as we now describe. In order to minimize the problem in \cref{eq:min}, we follow a proximal gradient descent approach by taking the gradient of the differentiable function $f(\x)$ and the proximal of the regularization function, which we define as $\mathcal P_R$, allowing for potentially non-smooth functions $R(\x)$. 
This algorithm produces the iterates given by:
\begin{align}\label{eq:pgd_iters}
    \x_{k+1} = \mathcal P_{R}(\x_k - \alpha_k \nabla f(\x_k)),
\end{align}
where $\alpha_k \in \mathbb{R}$ is the step size at the $k$-th iteration. 
The proximal operator of a function $R$ is formally given by
\begin{align}
    \mathcal P_{R}(\x) = \arg \min_\z \left( R(\z) + \frac{1}{2} \|\x-\z\|^2_2 \right),
\end{align}
thus providing estimates that are close to a given point while having a small value of the regularizer, $R(\x)$.
To leverage the advantage of deep neural networks for learning powerful priors from large amounts of data, herein we parameterize the proximal operator $\mathcal P_R$ by a deep neural network $G_\theta$, so that the iterates in \cref{eq:pgd_iters} become
\begin{align}
    x_{k+1} & = G_\theta(x_k - \alpha_k \nabla f(x_k)) \coloneqq S_\theta(x_k;y)
    \label{eq:Gtheta}
\end{align}
where we succinctly denote by $\theta$ all trainable parameters of the neural network. As we will shortly show, the model $G_\theta$ will learn the prior distribution of STI images implicitly by approximating the proximal operator for the regularizer. Importantly, this parameterization decouples the learned prior from the forward operator, allowing it to be applied with a variety of forward models. Indeed, a new forward model can be naturally incorporated in our algorithm by adjusting the forward function in Eq. \ref{eq:Gtheta} (i.e., by modifying the function $f(\x)$), without any modifications to $G_\theta$. In the context of STI, this will allow our model to deal with an arbitrary number of measurements at arbitrary head orientations without any retraining or adaptation of the network.

In this way, given an initial estimation $\hat{\x}_0$ and a trained model $G_{\hat \theta}$, the solution to the inverse problem is given by iteratively applying the following updates with the learned proximal operator:
\begin{align}
    \hat{\x} = (S_{\hat \theta} \circ \ldots \circ S_{\hat \theta})(\hat{\x}_0;\y) = S_{\hat \theta}^{(K)} (\hat{\x_0};\y) \coloneqq \phi_{\hat \theta}(\y),
\end{align}
where $K$ is the number of iterations and $\circ$ denotes the composition of functions.

In order to obtain a good model $G_{\hat \theta}$, we leverage a supervised learning framework. To this end, denote by $\mathcal{X}$ the distribution of susceptibility tensor images of human brains, and let $p(\mathrm Y \mid \x)$ be the distribution of the measurements $\y$ for a given tensor image. We then seek for model parameters, $\theta$, that minimize the expected loss over susceptibility tensors and their measurements, that is
\begin{align}
    \min_{\theta} \underset{\x \sim \mathcal{X}}{\mathbb{E}}~ \underset{\y \sim p(\mathrm Y\mid\x)}{\mathbb{E}} L(\phi_\theta(\y),\x).
    \label{eq:exp_loss}
\end{align}
The loss $L$ penalizes large differences between $\x$ and their estimates, $\hat\x=\phi_\theta(\y)$. In this work, we employ the $\ell_1$ norm to this end, i.e. $L(\phi_\theta(\y),\x)=\|\phi_\theta(\y)-\x\|_1$, to avoid the over-smoothing effect typically produced by the $\ell_2$ loss.
In practice, the distributions above are unknown and thus training is done by minimizing an empirical version of the  expected loss in \cref{eq:exp_loss}. Given a set of training samples of ground truth STI (were they known) and their corresponding measurements, $\{\x_i, \y_i\}_{i=1}^N$, we minimize this loss over training samples via the following optimization problem:
\begin{align}
    \hat \theta = \arg \min_\theta \frac{1}{N} \sum_{i=1}^N \|\phi_\theta(\y_i) - \x_i\|_1.
\end{align}
While this formulation can provide good learned predictors, ground-truth STI samples $\x$ cannot be easily obtained. To resolve this issue, we simulate this distribution by generating realistic brain phantoms using a combination of in vivo QSM and DTI measurements collected from human subjects. We can thus sample the measurements $\y_i$ for a given tensor image $\x_i$ with the forward model specified above in \cref{eq:forward}. We now move on to the generation of this training data.

\subsection{Training Data Generation}
\label{sec:method-phantom}
One major challenge of applying the learned proximal network for STI is the lack of ground-truth data. As stated above, none of the existing methods can provide accurate STI estimation for human in vivo (unlike QSM) and, as a result, it is not feasible to acquire high-quality ground-truth STI from in vivo human subjects, even with large amounts of sampling.
To tackle this issue, and to enable supervised learning with measurement-source image pairs, we resort to the available QSM and DTI datasets of in vivo human subjects for generating realistic brain phantoms for STI. In particular, multiple orientation QSM (COSMOS) can provide high-quality estimates of mean magnetic susceptibility. On the other hand, the fiber direction and fractional anisotropy (FA) estimated in DTI can provide good estimates for the principal eigenvector and scaled susceptibility anisotropy. Therefore, we can synthesize computational STI phantoms from QSM-DTI images of the same subjects. Local field measurements are then obtained from these using the forward model in \cref{eq:forward}. The simulated measurements and the susceptibility tensor source pairs are used as training samples for the learned proximal network.

More precisely, denote the mean magnetic susceptibility (MMS) value at each voxel as $q \in \mathbb{R}$, as obtained from QSM, and the principal eigenvector of DTI as $\v_{D}\in \mathbb{R}^3$ and the DTI FA as $a_D \in \mathbb{R}$. The  magnetic susceptibility anisotropy (MSA), $a_S$, is modelled from a scaled DTI FA: $a_S = \gamma a_D$, where $\gamma$ is a scaling factor to map $a_D$ to the range of $a_S$. Further, the eigenvalues of susceptibility tensor at each voxel are given by solving the following system:
\begin{align}
    \begin{cases}
    (\lambda_1+\lambda_2+\lambda_3) / 3 = q \\
    \lambda_1 - (\lambda_2+\lambda_3) / 2 = a_S \\
    \lambda_2 - \lambda_3 = \Delta
    \end{cases}
    \label{eq:lambda}
\end{align}
where $\lambda_1 \geq \lambda_2 \geq \lambda_3$ are the eigenvalues of the tensor at each location. Above, $\Delta$ quantifies the difference between the two smaller eigenvalues, known as the tensor asymmetry, which is $0$ under the ideal cylindrical symmetry assumption for myelinated nerve fibers. In our model, we randomly sample $\Delta$ for each voxel from a small range $[0,\epsilon]$ to capture the deviation from cylindrical symmetry in real world settings. 

Moreover, we let the principal eigenvector $\v_1\in \mathbb{R}^3$ of the susceptibility tensor at each voxel to be $\v_1 = \v_D$, since the principal eigenvectors should align with white matter fiber direction in both STI and DTI. The remaining two eigenvectors, $\v_2$ and $\v_3$, are sampled randomly such that $\v_1$, $\v_2$, $\v_3$ are orthonormal.
In this way, the susceptibility tensor at each voxel is given by
\begin{align}
    \chib = \V\Lambdab \V^T
\end{align}
where $\Lambdab = \text{diag}(\lambda_1, \lambda_2, \lambda_3)$ and $\V = \begin{bmatrix}\v_1,\ \v_2,\ \v_3
\end{bmatrix}$. Finally, the local field maps are computed from computational phantoms via the STI forward model described in \cref{eq:forward}, with additive Gaussian variables $\e$ to account for measurement noise and discrepancies between the idealized STI forward model and the real-world imaging system.

More details on the network architecture, training protocol, phantom generation and other implementation details can be found in the Supplementary Material.

\subsection{Data and competing methods}
Previously published GRE data from 8 healthy young adults (1 Female, mean(std) age of 32(3) years) were used. Data were from 3 subjects scanned at 3T (Philips, Achieva) with $(1.5~ \text{mm})^3$ isotropic resolution (data from \cite{li2014mean})and 5 subjects scanned at 7T (Philips, Achieva), with $(1~ \text{mm})^3$ isotropic resolution, reconstructed to $0.98\times0.98\times1~ \text{mm}^3$ (data from \cite{li2012mapping}). A varying number of 6 to 12 head orientations were sampled for each subject scanned at 3T, and 4 head orientations were sampled for each subject scanned at 7T. Phase preprocessing includes phase unwrapping using a Laplacian-based method \citep{schofield2003fast,li2011quantitative} and background field removal using the sophisticated harmonic artifact reduction on phase data with variable kernel size (VSHARP) method \citep{schweser2011quantitative,wu2012whole,wu2012fast} and maximum kernel radius of 4.5 mm for the 3T data (as in \cite{li2014mean}). For the 7T data, best-path based unwrapping \citep{abdul2007fast} and LBV+VSHARP methods \citep{zhou2014background} with maximum kernel radius of 8 mm were used to calculate the local field maps. Multiple orientation QSM images were reconstructed from all available GRE measurements using COSMOS. We also used previously published DTI data from these same subjects at 3T with $(2.2~\text{mm})^3$ isotropic resolution (data from \cite{li2014mean,li2012mapping}), reconstructed and interpolated to the same resolution as the GRE measurements of each subject. 

For data simulation, STI phantoms were generated for each subject using the proposed pipeline in \cref{sec:method-phantom} and local field measurements from different head orientations were simulated from the generated phantoms. To imitate the limitation on rotation angle from the head coil in real situation, the simulated rotation angles were chosen to be uniformly distributed in a range of $25^{\circ}$ with respect to the main magnetic field $B_0$. Gaussian noise at signal-to-noise-ratio (SNR) of 10dB (10:1) was added to all simulated measurements.

Experiments were performed by 5-fold cross validation, each time dividing the 8 subjects into 5, 1 and 2 subjects for network training, validation and testing, respectively. The testing data always included one low-resolution data acquired at 3T and one high-resolution data acquired at 7T.

In this work we compare with the following state-of-the-art methods for STI reconstructions: 1) STIimag \citep{li2017susceptibility}: an image-space-based STI method that solves \cref{eq:min} with a simple regularization setting the non-brain area to be close to zero; 2) Mean Magnetic Susceptibility regularized (MMSR) STI \citep{li2014mean}: a regularized STI method which limits anisotropy to white matter area and imposes morphology constraint on MMS; and 3) asymmetric STI (aSTI) \citep{cao2021asymmetric}, which fits an asymmetric tensor using a least squares solver and then decomposes it into symmetric and anti-symmetric components; and 4) aSTI+ \citep{shi2022regularized}: an improved version of aSTI with an isotropic constraint inside cerebrospinal fluid (CSF) and a morphology constraint based on MMS inside white matter. All methods were implemented in Matlab using a least square (LSQR) solver \citep{paige1982lsqr}. The convergence tolerance of the LSQR solver was optimized independently for each method using simulated measurements from 6 orientations to maximize their performance.

\subsection{Fiber Tractography}
To evaluate the performance of DeepSTI for white matter fiber tracking, we performed whole brain tractography and local fiber tracking through the corpus callosum (CC) using the STI results and compared these results to those using DTI. Whole-brain tracking of white matter fibers was performed using the FACT algorithm \citep{mori1999three}, and tracking was done using the MRtrix3 toolbox \citep{tournier2019mrtrix3}. Seeds were randomly sampled from inside the brain mask until a million tracks have been generated. The maximum length of track was set as 250 mm and the anisotropy cut-off threshold for terminating tracks was set as 0.01 ppm for STI MSA and 0.12 for DTI FA. Then, we performed tracking of local fiber bundles passing through the CC. To do this, a mask of CC was manually delineated on a mid-sagittal slice and all fibers passing through the mask were selected from the whole brain tracking result. 

We also performed tracking of a specific fiber bundle, i.e. the forceps major, the white matter fiber bundle that connects the left and right occipital lobes via the splenium of corpus callosum \citep{wakana2007reproducibility}. For this, we set the number of tracts for whole brain tracking to 10 million for both DTI and STI, and decreased the cutoff threshold for DTI FA to 0.08 to recover more tracks in the forceps major. Segmentation of the left and right occipital lobes was obtained using MRIstudio \citep{jiang2006dtistudio} (\url{http://www.mristudio.org}) and combined with the gray matter mask from FSL FAST \citep{zhang2001segmentation}. Tracks passing through the CC and both occipital regions were selected from the whole brain tracking as forceps major. Finally, the selected fibers were cleaned up based on \cite{yeatman2012tract} to remove outlier deviating tracks from the central bundle. 

\subsection{Error Metrics}
\label{sec:error-metrics}
The following error metrics were used to quantitatively evaluate the reconstruction accuracy of different STI methods.
First, to evaluate the accuracy of tensor image reconstruction, peak signal to noise ratio (PSNR) and structural similarity index measures (SSIM) were computed for the whole brain. SSIM was first computed for each tensor component separately and then averaged over all 6 components. To evaluate the accuracy of estimated fiber directions, we computed two angular metrics on PEV maps: eigenvector cosine similarity error (ECSE) and weighted-PSNR (wPSNR). ECSE is defined as one minus the cosine similarity between the estimated and ground-truth PEV, averaged over the anisotropic region (defined by voxels with ground-truth MSA $>$0.015 ppm). wPSNR is the PSNR of anisotropy-weighted PEV maps over the whole brain. The results are presented in \cref{sec:quantitative-results}.

\begin{figure*}[!t]
    \centering
    \includegraphics[trim=0 50 0 30, clip, width=\textwidth]{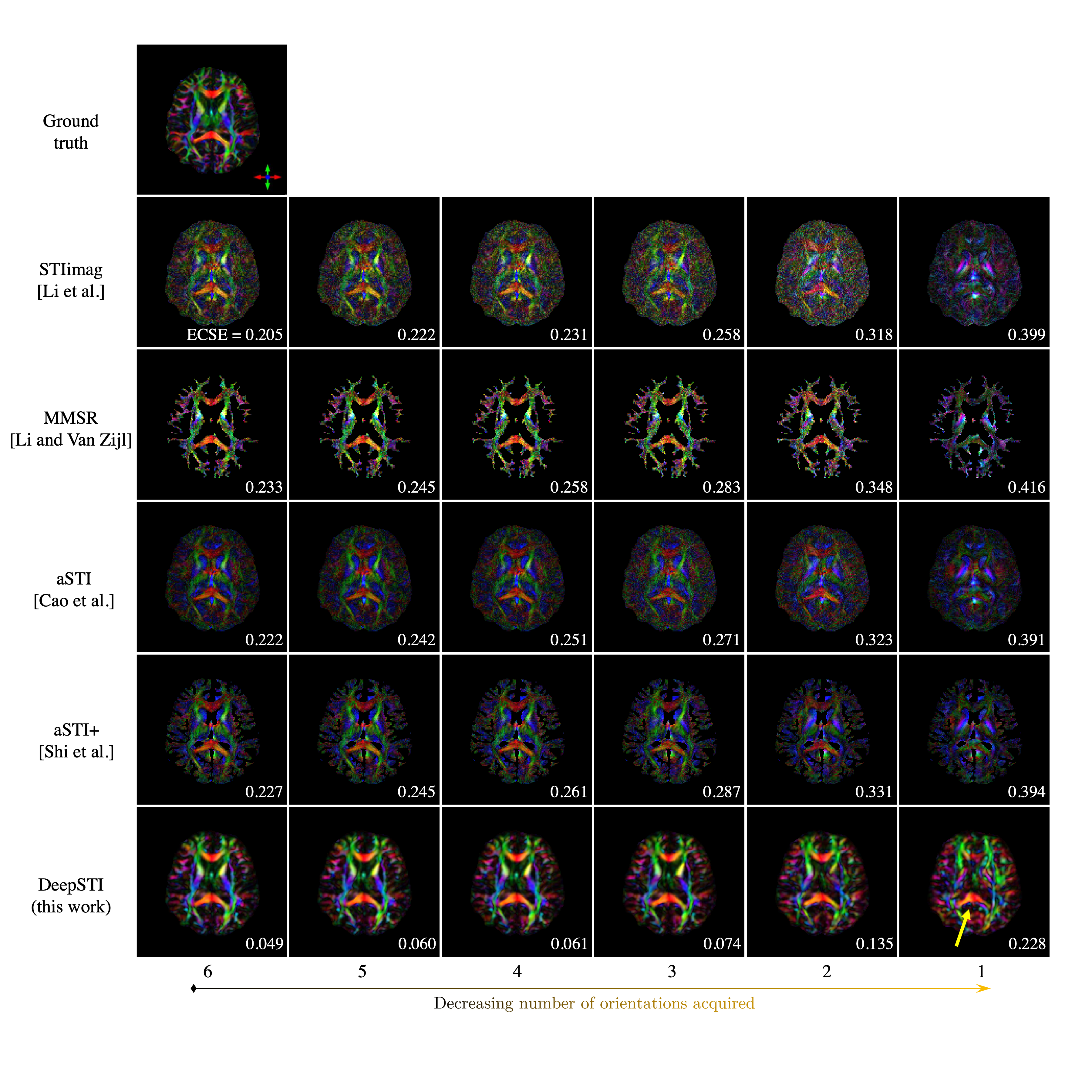}
    \caption{Magnetic susceptibility anisotropy (MSA) weighted PEV maps estimated from measurements at decreasing number of head orientations simulated from a brain phantom with 1.5 mm resolution. On the lower-right corner of each image shows the eigenvector cosine similarity error (ECSE) of estimated PEV as defined in \cref{sec:error-metrics}.}
    \label{fig:res-pev-sim}
\end{figure*}

\section{Experimental results}
\label{sec:results}

\subsection{Principal Eigenvector Estimation from Simulation}
\Cref{fig:res-pev-sim} presents the estimated principal eigenvector (PEV) maps obtained from different numbers of head orientations (from 6 to 1) using simulated measurements of a brain phantom with 1.5 mm isotropic resolution. All STI PEV maps are weighted by the magnetic susceptibility anisotropy (MSA) estimated by each respective method. Note that while the visual results are affected by both MSA and PEV estimation, the ECSE metric reflects only PEV estimation accuracy. A significant improvement can be observed in the results yielded by DeepSTI. Compared to the existing methods, more accurate PEV estimation is achieved by DeepSTI across all numbers of head orientations. Even when the number of orientations decreases from 6 to 2, DeepSTI continues to produce highly accurate results (as measured by ECSE) that match well with the ground-truth, while the estimations from other methods deteriorate drastically. In the extreme case of reconstruction from only one head orientation, DeepSTI can still recover some structures visible in the ground-truth PEV map, notably in the corpus callosum splenium (indicated by yellow arrow), while the other methods fail to produce any useful reconstructions. These results demonstrate the potential of accurate fiber direction mapping using STI when sampling a much smaller number of head orientations than previously reported.
 
\begin{figure*}[!t]
    \centering
    \includegraphics[trim=0 50 0 30, clip,
    width=\textwidth]{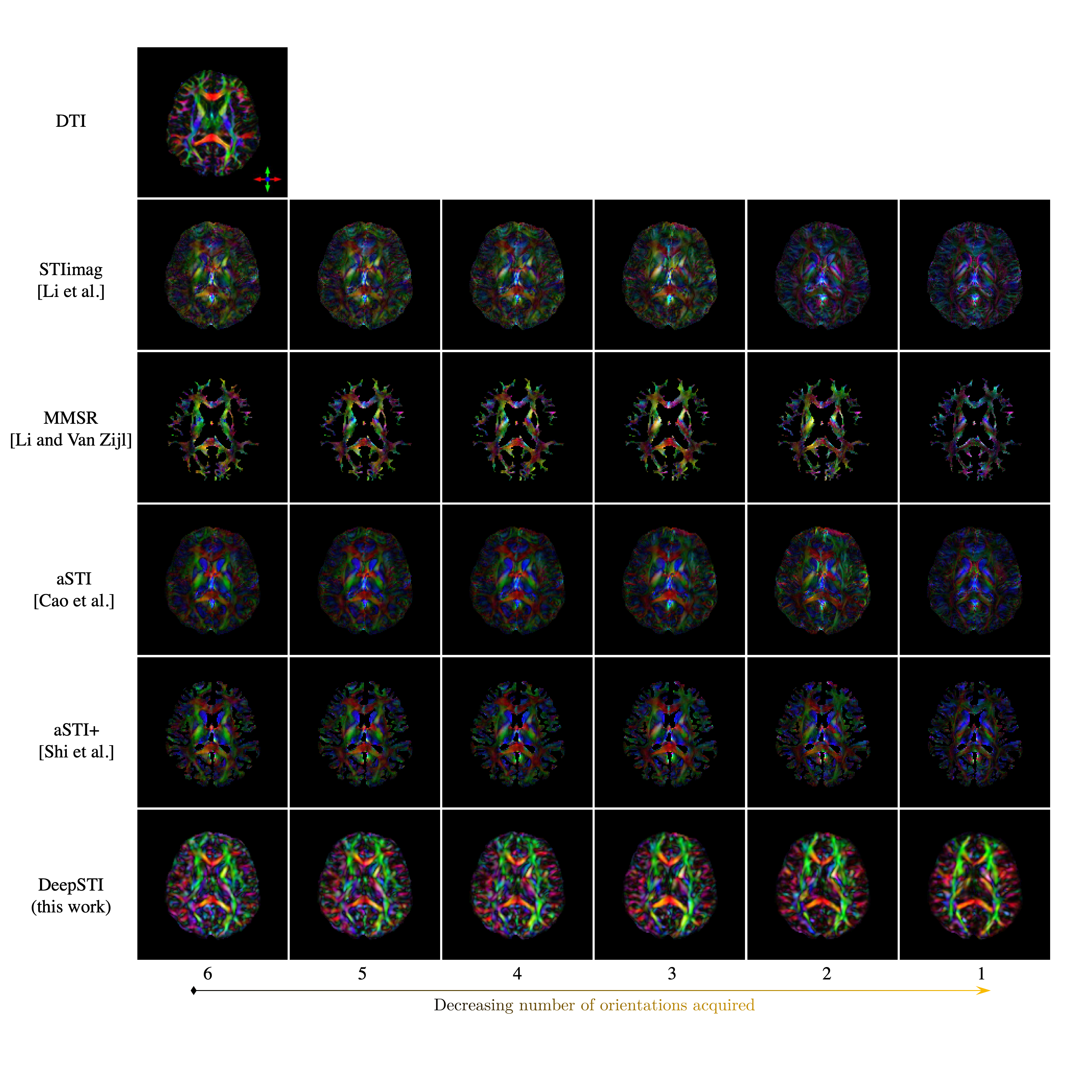}
    \caption{MSA-weighted PEV maps estimated from in vivo GRE measurements of a subject at 3T  with 1.5 mm isotropic resolution as a function of number of head orientations, with FA-weighted DTI PEV map as reference. DTI is acquired at 2.2 mm isotropic resolution and interpolated to 1.5 mm isotropic resolution. The detailed head orientations are specified in the appendix (\cref{tab:head-orientations}).}
    \label{fig:res-pev-real}
\end{figure*}

\subsection{Principal Eigenvector Estimation from in vivo Data}
We further evaluated DeepSTI using in vivo data. \Cref{fig:res-pev-real} shows the MSA-weighted STI PEV maps estimated from MR phase measurements acquired on a subject at 3T with 1.5 mm isotropic resolution, in comparison to the DTI PEV map. The DTI PEV map is weighted by the DTI FA, while all STI PEV maps are weighted by MSA estimated by each respective method. Note that the qualitative performance of all STI methods decreased as compared to the simulation, likely due to errors introduced in pre-processing steps such as co-registration, background field removal and potential MR phase/frequency contributions that are unaccounted for in the forward model, e.g. multi-compartment microstructure effects or chemical exchange etc. \citep{wharton2012fiber,sati2013micro,wharton2015effects}. Nonetheless, the results provided by DeepSTI still appear more anatomically coherent and better match the result of DTI visually. Note that with only one orientation, DeepSTI can still yield a reasonable estimation of the PEV map.

\subsection{Susceptibility Tensor Component Estimation}

\begin{figure*}[!t]
\centering
\subcaptionbox{Simulation results from computational phantom.
\label{fig:res-tensor-sim}}
{\includegraphics[trim={30 115 0 140},clip,width=\linewidth]{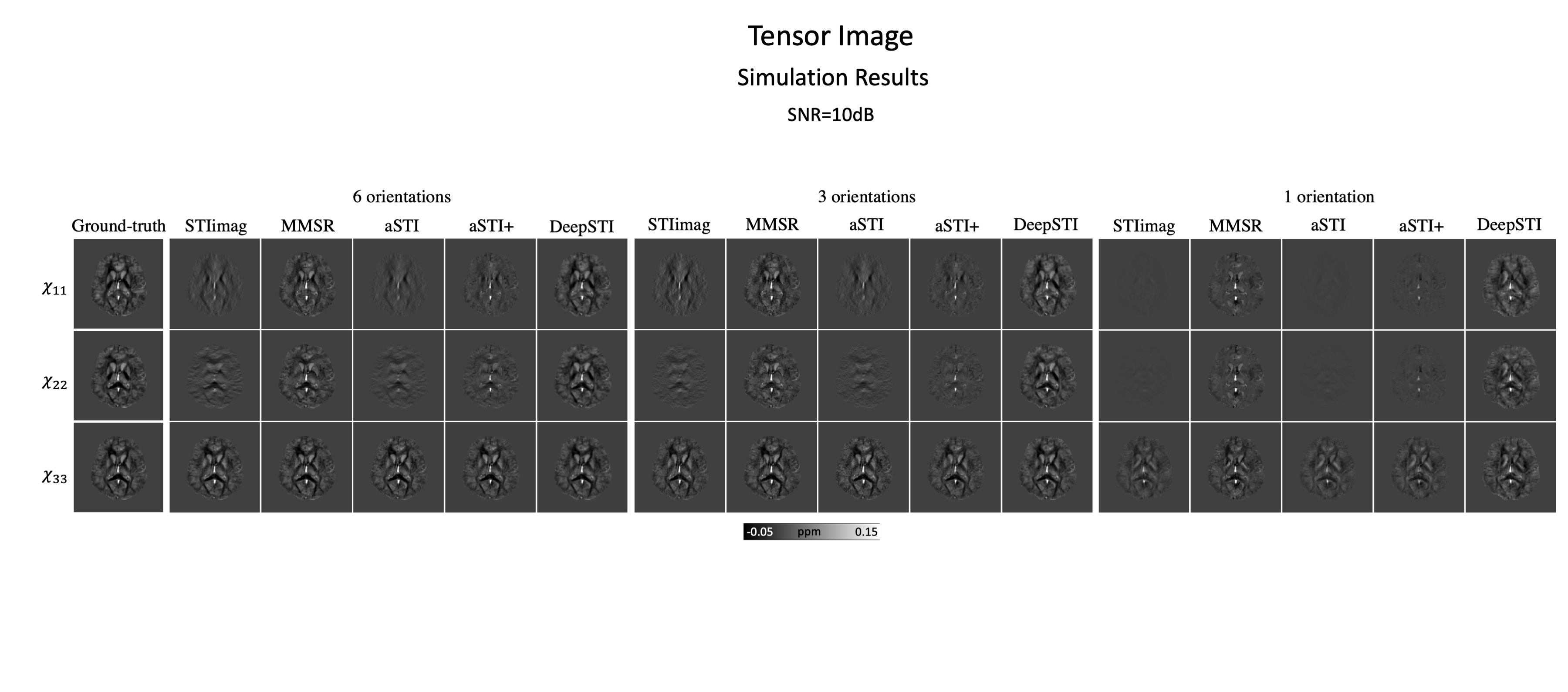}}
\subcaptionbox{Results from in vivo human measurements.
\label{fig:res-tensor-real}}
{\includegraphics[trim={25 115 5 140},clip,width=\linewidth]{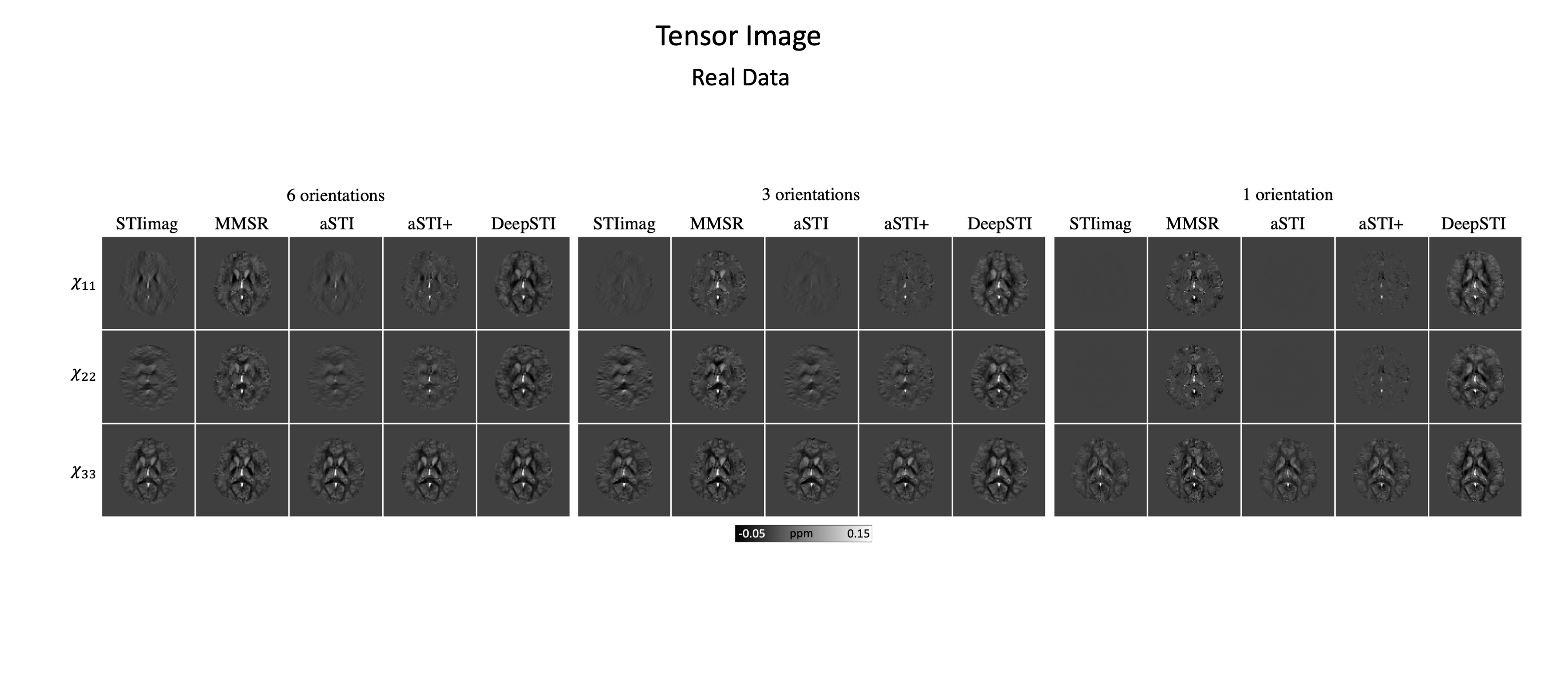}}
\caption{Diagonal elements of tensor images ($\chi_{11}, \chi_{22}, \chi_{33}$) reconstructed by different methods using (a) simulated and (b) in vivo phase measurements from different numbers of head orientations.}\label{fig:res-tensor}
\end{figure*}

We further visualize the susceptibility tensor component images reconstructed by different methods in  \cref{fig:res-tensor-sim} and \cref{fig:res-tensor-real} using the same 3T subject as shown in \cref{fig:res-pev-sim,fig:res-pev-real}. The three diagonal elements of the tensor are shown, defined in left-right ($\chi_{11}$), anterior-posterior ($\chi_{22}$) and superior-inferior ($\chi_{33}$) directions, respectively. \Cref{fig:res-tensor-sim} shows results from numerical simulations, along with the ground-truth tensor images from the brain phantom. It can be observed that with 3 head orientations, all three methods reconstructed $\chi_{33}$ accurately, but DeepSTI outperforms other methods for $\chi_{11}$ and $\chi_{22}$ estimations. With only one head orientation, DeepSTI produced reasonably good results for all three diagonal elements, while the performance of other methods had deteriorated substantially.  \Cref{fig:res-tensor-real} shows results from in vivo measurements, where DeepSTI again outperforms other methods for all numbers of measurements and degrades the least with decreasing number of head orientations.

\subsection{Tractography}
\label{sec:track-results}
\begin{figure*}[!h]
    \centering
    \includegraphics[trim=0 200 0 0,clip,width=\linewidth]{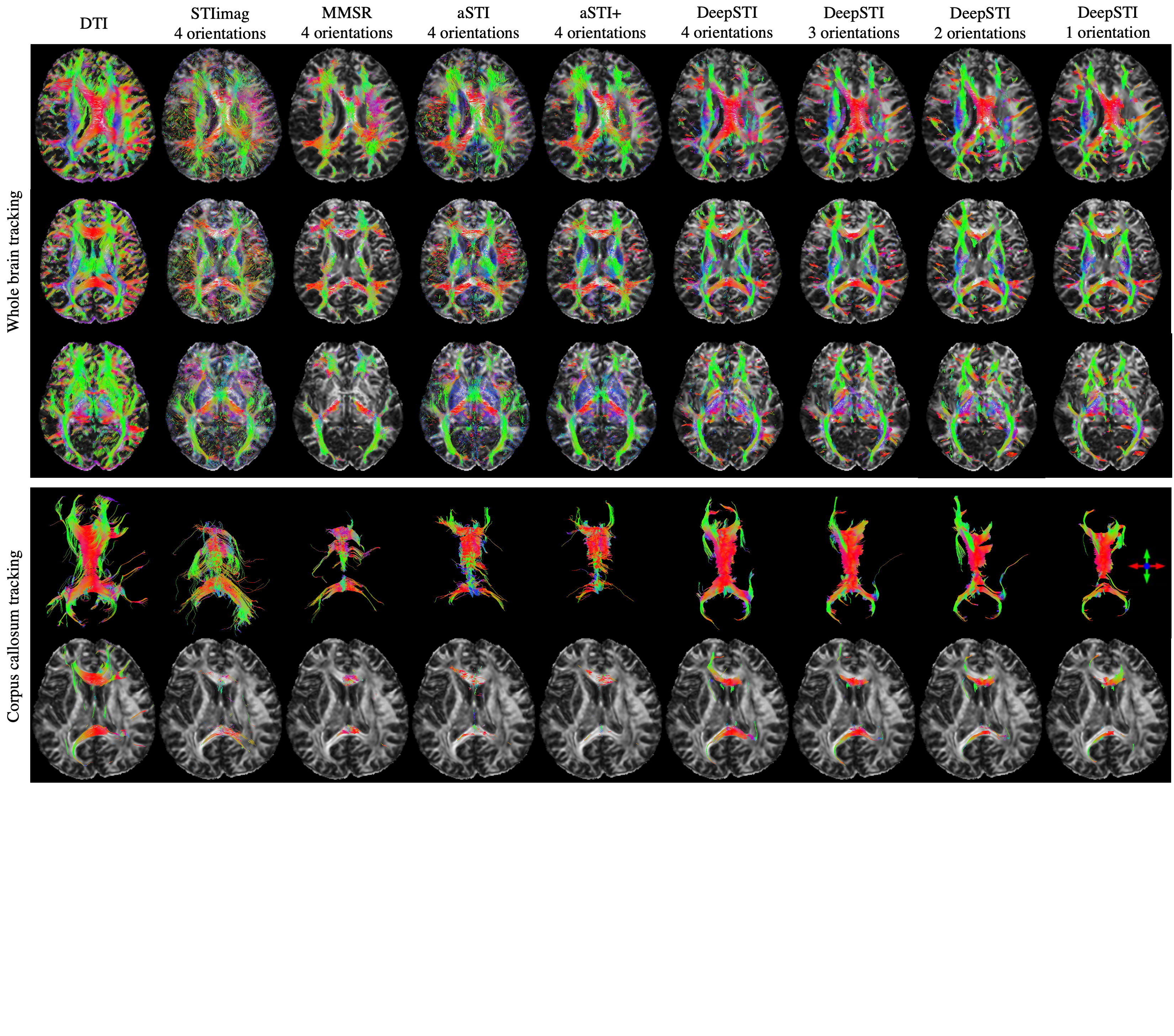}
    \caption{STI-based Fiber tractography results  from phase measurements for a subject at 7T with 0.98$\times$0.98$\times1 (mm)^3$ resolution. Rows 1-3: whole brain tracking results overlaid on a DTI-based FA map at different axial planes. Row 4: 3D volume rendering of neural fibers passing corpus callosum. Row 5: 2D visualization of neural fibers passing corpus callosum. Columns from left to right: DTI, STIimag's result from 4 orientations, MMSR's result from 4 orientations, aSTI's result from 4 orientations, aSTI+'s result from 4 orientations, DeepSTI's results from 4, 3, 2 and 1 orientations. No FA or MSA weighting is used in visualizing the tractography results.}
    \label{fig:res-track-real-7T}
\end{figure*}

\Cref{fig:res-track-real-7T} depicts fiber tracking results achieved by different methods using in vivo data acquired at 7T with 0.98$\times$0.98$\times$1 mm resolution. With 4 head orientations, DeepSTI provides more complete fiber reconstructions as compared to other STI methods. Even with fewer head orientations, DeepSTI can still recover a good portion of major fiber pathways in whole brain tracking (top row of Fig. 5). Notably, the fibers passing through CC connecting the left and right hemispheres can still be nicely recovered even from only one head orientation (middle and bottom rows of \cref{fig:res-track-real-7T}). Results for a 3T subject are included in the Appendix (\cref{fig:res-track-real-3T}).

\begin{figure}[!h]
    \centering
    \includegraphics[trim={0 5.2in 3.5in 0},clip,width=3.5in]{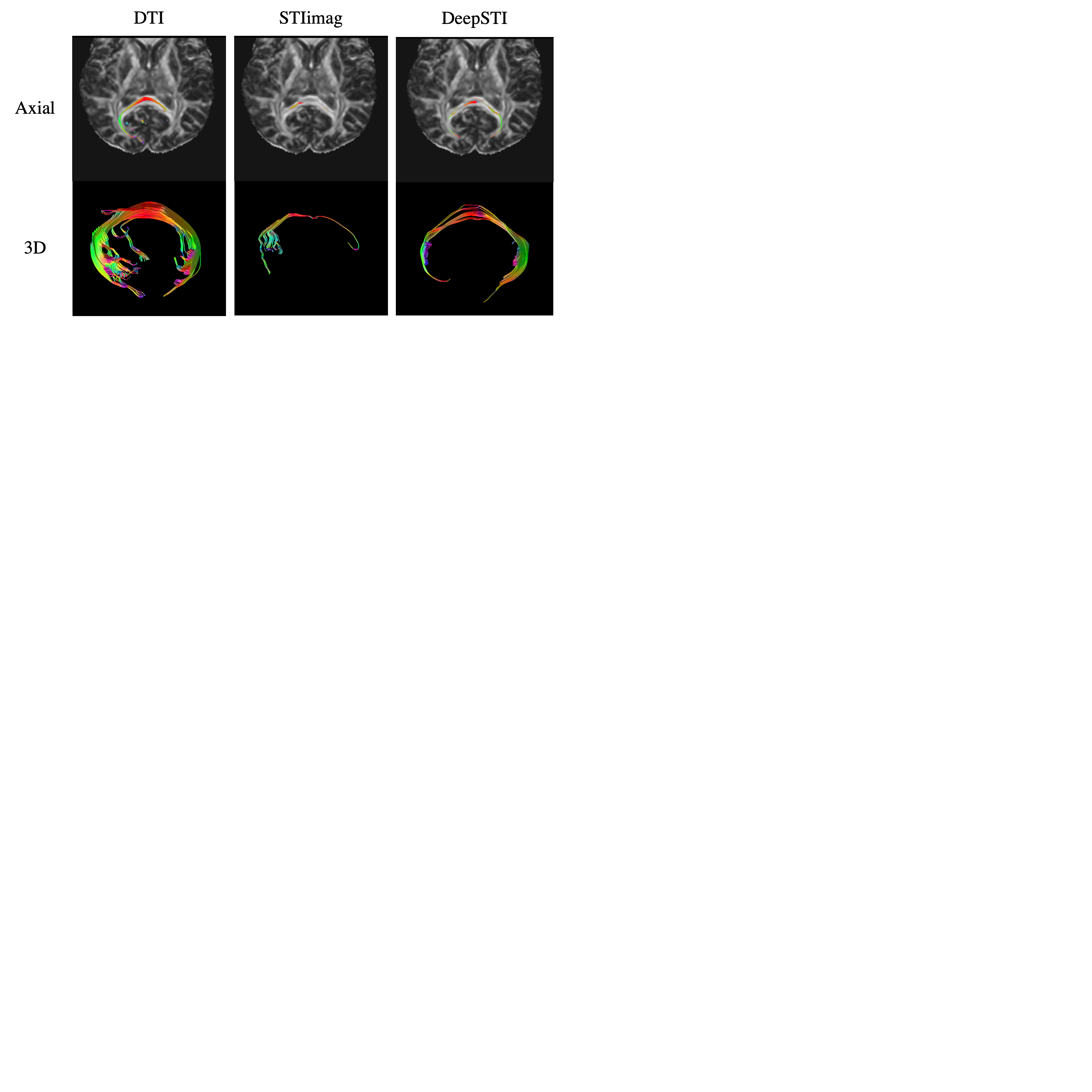}
    \caption{Tracking of callosal fiber bundle that connects the occipital lobes via the splenium of corpus callosum (forceps major) from measurements of a subject at 7T with 0.98$\times$0.98$\times1 (mm)^3$ resolution. From left to right: reference DTI reconstruction (acquired at 2.2 mm isotropic resolution and interpolated to 0.98$\times$0.98$\times1 (mm)^3$ resolution), STI reconstruction from STIimag using four orientations, STI reconstruction from DeepSTI using four orientations. Note that MMSR, aSTI and aSTI+ did not recover any tracts that satisfy the anatomical constraint of the forceps major.}
    \label{fig:res-track-real-occipital}
\end{figure}

\cref{fig:res-track-real-occipital} depicts tractography results for the forceps major using the same 7T dataset as in \cref{fig:res-track-real-7T}.  In this task, most existing STI methods (including MMSR, aSTI and aSTI+) failed to reconstruct the forceps major, yielding no tracts that satisfy the given anatomical constraint. STIimag was able to produce a small portion of tracts, but largely incomplete. DeepSTI yielded the most coherent and complete result among all STI methods, with a complete C-shape structure.

\subsection{Variance Analysis across Different Angles}
\begin{figure}[!t]
    \centering
    \includegraphics[trim = 10 0 10 80, clip, width=.49\textwidth]{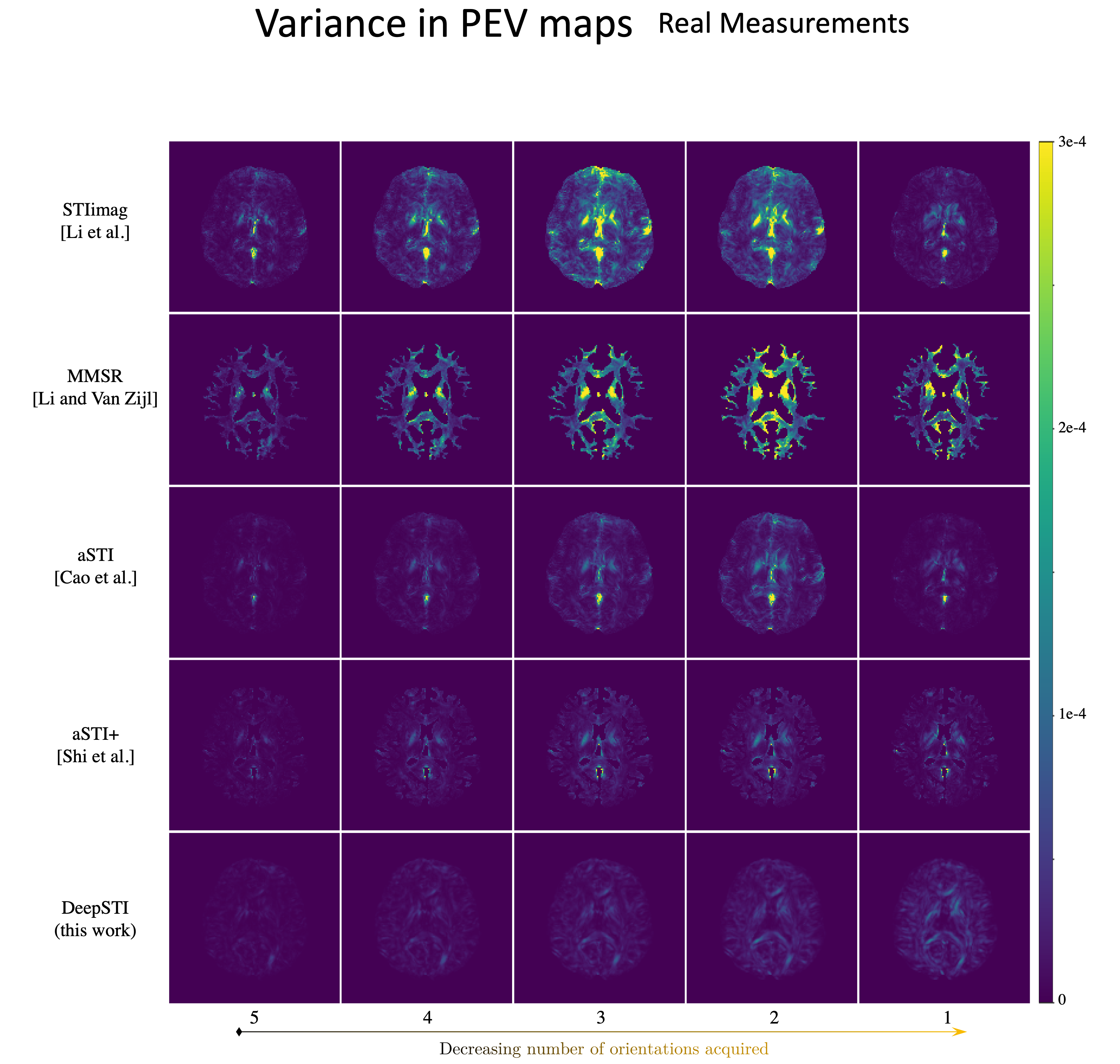}
    \caption{Variance in MSA-weighted PEV maps estimated from different observing angles using a 3T human in vivo measurements. A total of six head orientations were acquired from this subject. At each number of head orientations, all possible combinations of angles (6,15,20,15 and 6 combinations for 5,4,3,2 and 1 orientations, respectively) were tested.}
    \label{fig:res-var-real}
\end{figure}

Since DeepSTI can reconstruct the underlying susceptibility tensor using MR phase measurements obtained with different head orientations, it is important to understand the variation in the estimated tensor at various angle combinations for the reduced number of measurements. To this end, we further computed the variance in PEV maps estimated from different observation angles for each STI method. \Cref{fig:res-var-real} shows the result from a 3T data, where 6 head orientations in total were acquired. For each number of orientations less than 6, we computed STI reconstruction from all possible combinations of angles (6, 15, 20, 15 and 6 combinations for 5, 4, 3, 2 and 1 orientations out of 6, respectively) and presented the variance in the modulated PEV maps. As can be observed, the variance yielded by DeepSTI  is smaller than for other methods, and it increases only moderately with a reduction in the number of orientations. 

\subsection{Quantitative Results}
\label{sec:quantitative-results}
We quantitatively compare different methods by computing errors of the tensor reconstruction and PEV estimation from simulated measurements of brain phantoms---where ground-truth STI tensors exist---using the error metrics defined in \cref{sec:error-metrics}. To account for the variation in rotation angle, three different sets of head orientations were tested for each subject at each number of orientations. All orientations were uniformly sampled from within $25^\circ$ with respect to the main magnetic field. Results are summarized in Table \ref{tab:sim}. DeepSTI outperforms the other methods in all metrics with 6 and 3 orientations and in SSIM, ECSE and wPSNR with 1 orientation, which quantitatively verifies the improvement seen visually in the qualitative results.

\begin{table*}[!h]
    \centering
    \begin{tabular}{cccccc}
        \hline\hline
         Number of orientations & Method & PSNR & SSIM & ECSE & wPSNR \\
         \hline
         \multirow{5}{*}{6}     & STIimag  & 40.26(1.16) & 0.88(0.01) & 0.25(0.03) & 15.78(1.24) \\
                       & MMSR      & 44.14(0.82) & 0.91(0.01) & 0.26(0.03) & 16.98(0.70) \\
                       & aSTI   & 40.38(1.20) &	0.90(0.01) &	0.28(0.03) &	17.46(0.81) \\
                       & aSTI+ & 41.00(1.26) & 0.90(0.01) & 0.28(0.03) & 17.65(0.91) \\
                       & DeepSTI (this work)     & 47.52(0.80) & 0.96(0.01) & 0.08(0.01) & 23.20(1.00) \\
         \hline
         \multirow{5}{*}{3}     & STIimag & 39.24(1.24) & 0.87(0.01) & 0.30(0.03) & 15.07(1.51) \\
                       & MMSR      & 42.76(0.98) & 0.90(0.01) & 0.32(0.03) & 15.94(0.83) \\
                       & aSTI & 39.64(1.29) &	0.89(0.01) &	0.31(0.03) &	17.27(0.86) \\
                       & aSTI+ & 39.99(1.25) & 0.89(0.01) & 0.32(0.03) & 17.64(0.81) \\
                       & DeepSTI      & 44.88(1.19) & 0.95(0.01) & 0.12(0.02) & 21.97(1.00) \\
         \hline
         \multirow{5}{*}{1}     & STIimag & 38.27(1.13) & 0.87(0.01) & 0.38(0.02) & 16.92(0.81) \\
                       & MMSR      & 40.77(0.88) & 0.89(0.01) & 0.39(0.01) & 16.86(0.43) \\
                       & aSTI & 38.35(1.15) &	0.89(0.01) &	0.38(0.02) &	17.66(0.61) \\
                       & aSTI+ & 38.65(1.13) & 0.89(0.01) & 0.38(0.02) & 17.60(0.64) \\
                       & DeepSTI      &  40.13(1.33) & 0.91(0.01) & 0.25(0.03) & 19.61(0.76) \\
        \hline
        \hline
    \end{tabular}
    \caption{Quantitative metrics of STI reconstruction using simulated measurements from computational brain phantom, comparing DeepSTI to STIimag \citep{li2017susceptibility}, MMSR \citep{li2014mean}, aSTI \citep{cao2021asymmetric} and aSTI+ \citep{shi2022regularized}. Results were obtained from all 8 subjects using cross validation. PSNR, SSIM and wPSNR were computed for the whole brain and ECSE was computed from the anisotropic regions. Detailed definitions of the metrics are provided in \cref{sec:error-metrics}. Numbers in brackets denote standard deviation.}
    \label{tab:sim}
\end{table*}

\subsection{Multiple Sclerosis Patients}
\begin{figure*}
    \centering
    \includegraphics[trim=0 320 0 0, clip, width=.9\textwidth]{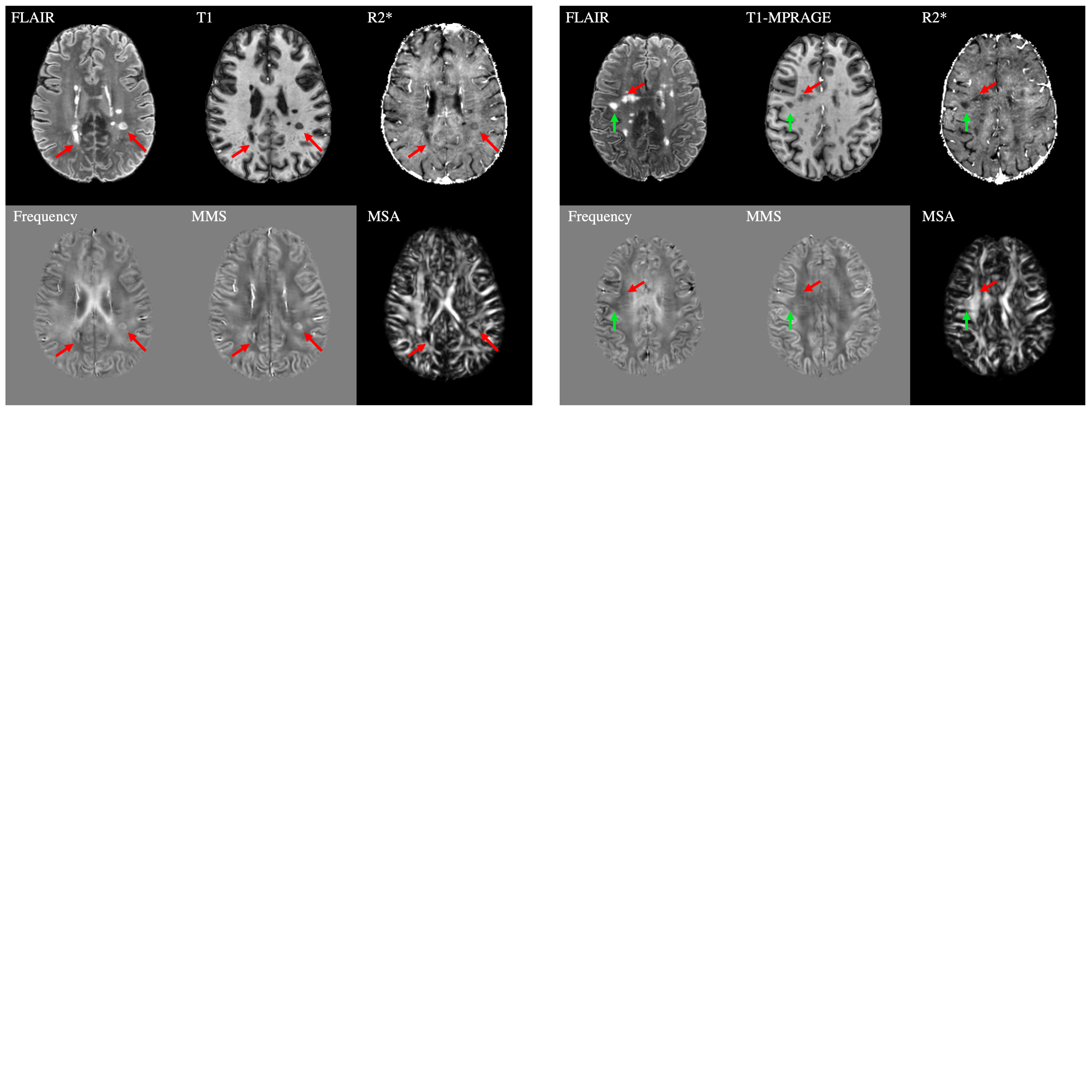}
    \caption{DeepSTI predicted mean magnetic susceptibility (MMS) and magnetic susceptibility anisotropy (MSA) from MR phase measured at one head orientation in two multiple sclerosis patients .}
    \label{fig:res-MS}
\end{figure*}

To further examine the generalization ability of DeepSTI to patients with neurological diseases, we tested DeepSTI using previously published GRE data from two multiple sclerosis (MS) patients (7T data from \cite{li2016magnetic}). Fig. \ref{fig:res-MS} depicts the MMS and MSA maps calculated from DeepSTI using only one head orientation for each patient. For the first patient (Fig. \ref{fig:res-MS}.a), DeepSTI predicted increased MMS and decreased MSA in two example lesions that are hyperintense in local frequency (and QSM as noted in \cite{li2016magnetic}) and hypointense in R2* (indicating major demyelination), as indicated by red arrows. For the second patient (Fig. \ref{fig:res-MS}.b), two example lesions that appear invisible (isointense) in local frequency (and QSM as noted in \cite{li2016magnetic}), but hypointense in R2* (indicating potential loss of both iron and myelin), and visible in T1 and FLAIR were selected. DeepSTI's prediction presents isointense in MMS and shows slightly decreased MSA (green arrow) in one lesion, and shows isointense MSA (red arrow) in the other lesion.

\subsection{Ablation Studies and Limitations}
To better understand the contribution of the learned prior to the reconstructed MSA of DeepSTI, we tested DeepSTI on simulated phase generated from a \emph{fully isotropic} STI brain phantom (manually set MSA=0 across the whole brain). Furthermore, to mimic the more realistic case of partially isotropic regions caused by lesions, we also tested our method on a local isotropic phantom where two cubic isotropic regions of size $(10~\text{voxel})^3$ were inserted in the original anisotropic phantom (\cref{fig:res-iso-local}).

\begin{figure*}[!h]
\centering
\subcaptionbox{
\label{fig:res-iso-hist}}
{\includegraphics[trim={140 230 130 40},clip,width=3in]{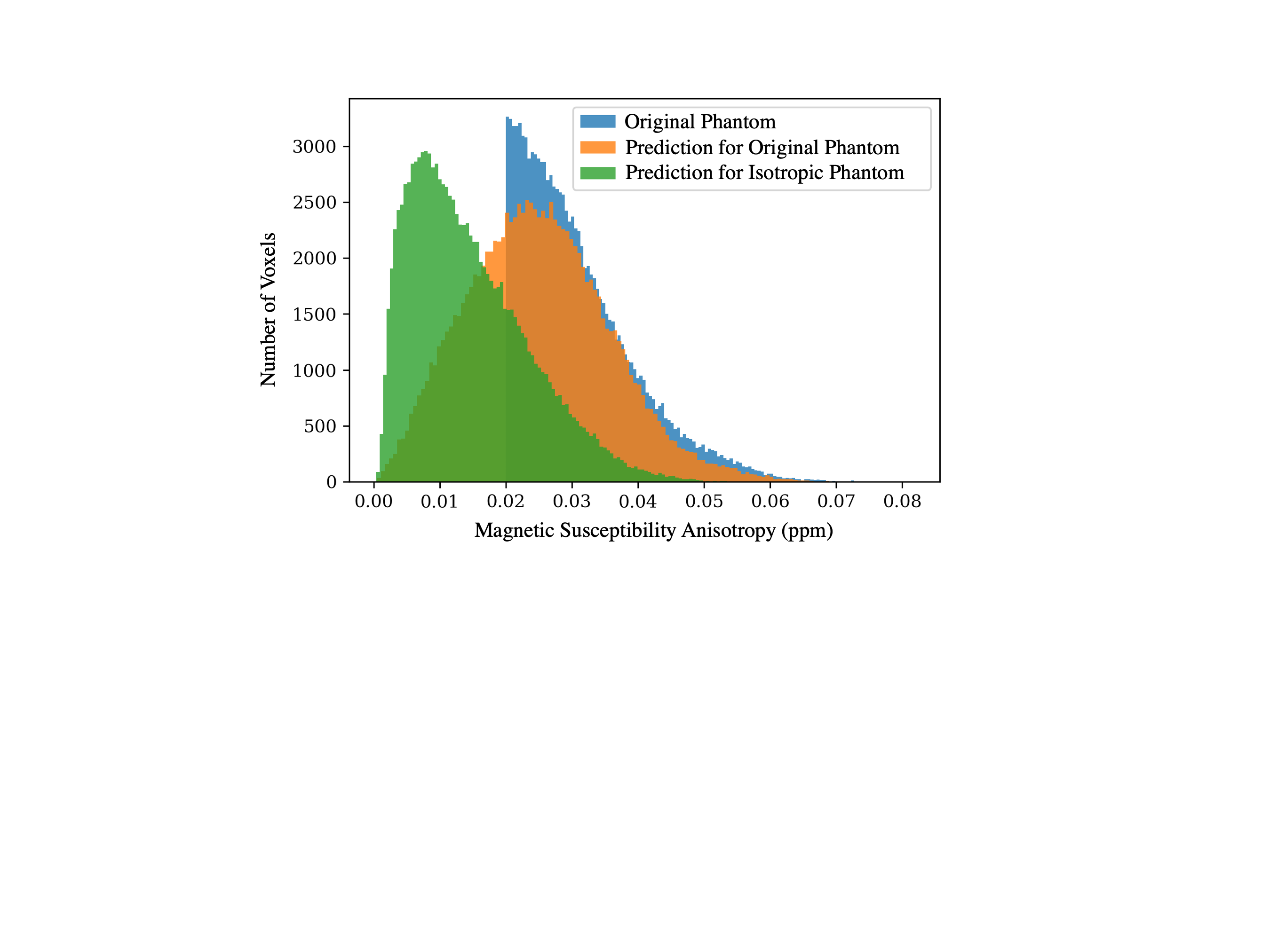}}
\subcaptionbox{
\label{fig:res-iso-full}}
{\includegraphics[trim={120 240 80 70},clip,width=4in]{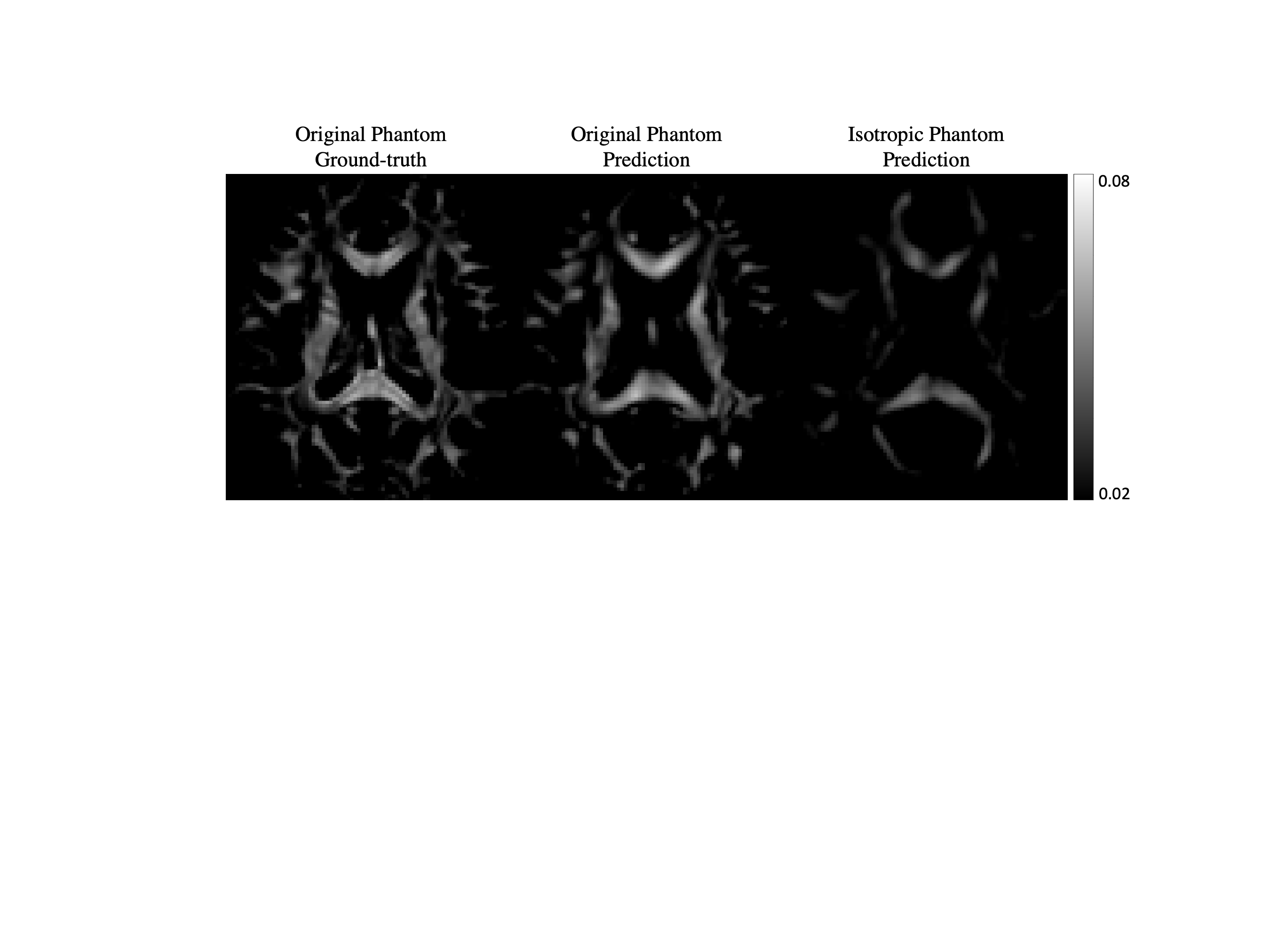}}
\caption{DeepSTI predicted MSA for a fully isotropic phantom. (a) Histogram of MSA values in major fiber regions defined by voxels with ground-truth anisotropy $>0.02 ppm $. (b) DeepSTI predicted MSA for the original anisotropic phantom and the isotropic phantom.}
\label{fig:res-iso-hist+full}
\end{figure*}

\begin{figure}[!h]
\centering
\includegraphics[trim={0 170 250 0},clip,width=3in]{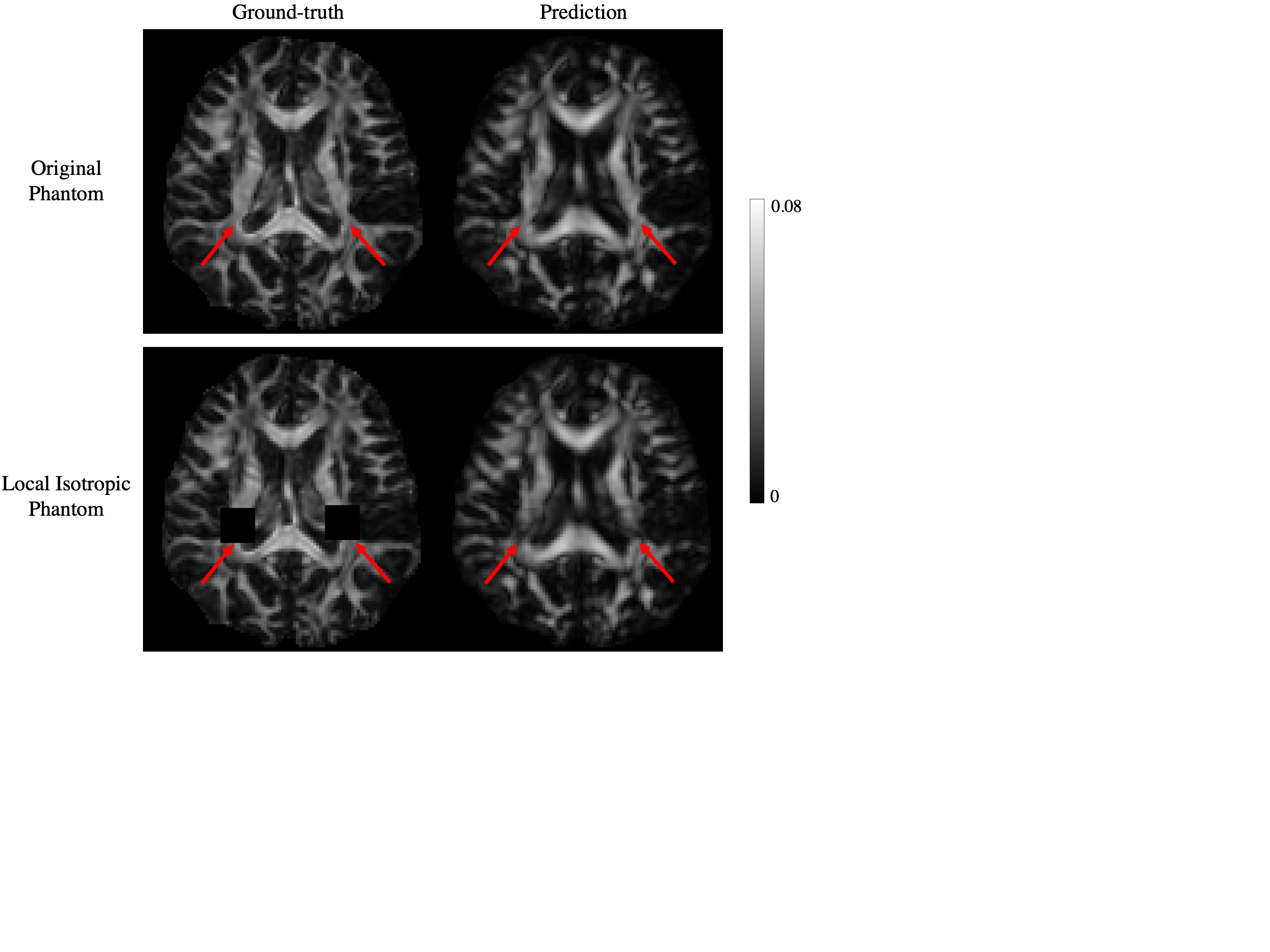}
\caption{DeepSTI prediction for a locally isotropic brain phantom, in comparison with the corresponding prediction for the original anisotropic phantom. Top row: ground-truth MSA (left) and DeepSTI-predicted MSA (right) for the original anisotropic phantom. Bottom row: ground-truth MSA (left) and DeepSTI-predicted MSA (right) for the locally isotropic phantom with two manually selected isotropic regions (indicated by red arrows).}
\label{fig:res-iso-local}
\end{figure}

\cref{fig:res-iso-hist} shows the histogram of MSA values predicted by DeepSTI in major fiber regions (defined by voxels with ground-truth MSA $>$0.02 ppm) for the original anisotropic phantom and the fully isotropic phantom, while \cref{fig:res-iso-full} shows corresponding MSA maps. Although DeepSTI naturally does not predict exactly zero MSA for the isotropic phantom, it is clear that the MSA values predicted by DeepSTI are significantly smaller in the isotropic phantom than in the original phantom. \cref{fig:res-iso-local} depicts the MSA maps predicted by DeepSTI for the local isotropic phantom, where a notable decrease in anisotropy can be observed in DeepSTI's prediction for both of the manually-defined isotropic local regions (indicated by red arrows).

\section{Discussion}
\label{sec:discussion}
In vivo human STI has long been hampered by the requirement for acquisition at multiple head orientations and the difficulty in reliable dipole inversion from noisy phase measurements. In this paper, we proposed a new data-driven method for STI dipole inversion that combines the benefits of modern deep learning approaches for learning powerful priors and the flexibility of classic iterative gradient-based algorithms for handling variable forward models.
The results show that DeepSTI provides estimations that are significantly more accurate (in the case of brain phantom data with ground truth) or more anatomically consistent in the case of real measurements, even though the model was trained on synthetic data, emphasizing its generalization ability. Notably, DeepSTI is able to manage diverse inputs at different resolutions, with different numbers of orientations, and allows for STI reconstructions with fewer phase measurements (less than 6).

While DeepSTI shares similarities to the ideas proposed in \citep{lai2020learned}, there are several differences. First, instead of training on estimated ground-truth images from COSMOS, we developed a phantom-based training scheme that addresses the issue of lacking ground-truth samples from real measurements in STI. Moreover, a new network architecture based on a Residual Symmetric U-Net \citep{ronneberger2015u,lee2017superhuman} was used to learn the proximal operator instead of the Wide ResNet \citep{zagoruyko2016wide}. This led to faster and more stable training as well as superior results in the reconstruction of the tensor image and estimation of white matter fiber directions, especially in the case of very few (e.g. single) head orientations. Finally, DeepSTI leverages a new multi-resolution training strategy that allows the model to handle measurements at different resolutions without the need for any ad-hoc re-sampling or re-training. Being able to handle data at different resolutions and different numbers of orientation sampling is central to STI reconstructions, where acquisition protocols between subjects and studies often differ.

The tensor image results in \cref{fig:res-tensor} reveal that the existing methods are far less accurate in estimating left-right ($\chi_{11}$) and anterior-posterior ($\chi_{22}$) elements than the superior-inferior ($\chi_{33}$) elements, which is consistent with results reported in \cite{bao2021diffusion}. This is not unexpected because the main magnetic field is in the superior-inferior direction and the head rotation angles are limited to a relatively narrow range ($<25^{\circ}$) around that. Note that in the lab frame of reference that aligns with the main field, only ($\chi_{33}$) and some off-diagonal elements ($\chi_{13}$ and $\chi_{23}$) contribute to the field shift and the MR phase measurement \citep{li2012mapping,milovic20202016}. However, DeepSTI was able to produce highly accurate estimations for all three diagonal elements, explaining its better performance in estimating the principal eigenvectors that are used to indicate white matter fiber directions as shown in \cref{fig:res-pev-sim} and \cref{fig:res-pev-real}.

Fig. \ref{fig:res-MS} demonstrates the result of our method on two MS patients. While different disease presentations and their neurobiological bases in STI remain to be analyzed in the future, this simple case-study demonstrates that the developed method has the potential to generalize to different clinical settings, e.g. for investigating lesion iron and myelin changes through the measures of MMS and MSA in MS with a single phase measurement. Comparing these results to the positive and negative susceptibility sources using recent chi-separation methods may also be helpful to better interpret the DeepSTI results in MS lesions \citep{chen2021decompose,shin2021chi}. Moreover, these results might provide new insights on tissue microstructure changes, such as the manifestation of reduced anisotropy on some MS lesions. We believe this illustrates the potential of DeepSTI to study similar neurological disorders in the future.

\cref{fig:res-iso-hist+full} presents the results on an unrealistic but interesting case, where reconstruction of a completely isotropic brain phantom is attempted. Although the results indeed show decreased MSA in white matter areas compared to the anisotropic phantom, the predicted anisotropy values are not completely zero. This is expected, because the solution given by DeepSTI can be broadly regarded as a maximum a posteriori (MAP) estimate for the ill-posed STI reconstruction problem, maximizing the probability of the solution under the learned prior from data. A completely isotropic brain has probability zero in the distribution of human STI (and, by extension, in all the training data used to train the model) and thus a completely isotropic reconstruction should never be expected. Yet, this does not mean that regions of abnormal anisotropy, or areas that are mostly isotropic in white matter, could not be approximately reconstructed. On the contrary, this is demonstrated by the experiment with the locally-isotropic regions in \cref{fig:res-iso-local}, as well as by the results on the MS patients.

Lastly, there are several limitations and areas of improvement in our study. First, our evaluation is based on a small cohort of subjects. Cross validation was employed to alleviate this limitation, but a large-scale evaluation involving samples from a broader and larger population will allow for both better training and characterization of the generalization ability of our method. Second, the results from real GRE measurements from human subjects are encouraging and more anatomically feasible than other methods, but these still lack ground-truth for validation and error calculation. In future studies, animal models with available ground-truth images, e.g. from high quality, large number of orientation sampling with no orientation constraints \citep{chen2022resolve,gkotsoulias2022beyond},  could be utilized to allow for better error quantification and further improvement of DeepSTI in vivo. Third, a recent study \citep{cao2021asymmetric} demonstrated that an asymmetric STI formulation---which removed the symmetric constraint on susceptibility tensors---could help separate noise and artifacts from the underlying susceptibility tensor sources, leading to better reconstruction results. DeepSTI can be incorporated with such asymmetric formulation as well, and other potential improvements in the physical model of STI might further improve the prediction of susceptibility tensor using DeepSTI. Finally, while the obtained networks have been proven to be very useful in regularizing this challenging image problem, it is still unclear to us how to precisely, analytically characterize and analyze the prior implicitly parameterized by these models.
Further investigation is required to better understand these data-driven priors to increase interpretability and reliability.

\section{Conclusion}\label{sec:conclusion}
In this paper, we presented DeepSTI for STI dipole inversion with the goal of reducing the need for current cumbersome and time-consuming STI acquisition schemes that require many head orientations. By combining the benefits of modern deep learning approaches for learning powerful priors and the flexibility of classic iterative gradient-based algorithms for handling variable forward models, DeepSTI enabled accurate and anatomically coherent tensor image reconstructions with fewer head orientations. Experimental results on both simulation and in vivo human brain data demonstrate the superiority of DeepSTI compared to state-of-the-art methods for tensor image reconstruction, fiber direction estimation and tractography. Our results shed light on potential large-scale application of high-resolution STI on human in vivo for better understanding brain functions and neurological diseases.

\section*{Acknowledgments}
This research has been supported by NIH Grant P41EB031771, as well as by the Toffler Charitable Trust and by the Distinguished Graduate Student Fellows program of the KAVLI Neuroscience Discovery Institute.

\bibliographystyle{model2-names.bst}\biboptions{authoryear}
\bibliography{refs}

\newpage
\appendix

\onecolumn

\section{Implementation Details}
Our network architecture for proximal learning was implemented following the Residual Symmetric Unet \cite{lee2017superhuman}. It contains a contracting path with five downsampling layers and an expanding path with five upsampling layers. Downsampling is implemented by max-pooling, and upsampling is by transpose convolution. The basic module for feature extraction is a single convolution followed by a residual block. The residual block contains two concatenated convolution layers, each followed by group normalization and Exponential Linear Unit (ELU) activation. Skip connections are added between the same spatial scale of contracting and expanding paths and summation joining is used instead of concatenation.

For network training, we used the Adam optimizer \citep{kingma2014adam} with a batch size of two. Patches of size $64^3$ were extracted with a stride of $32$ as training samples. Training was run for a total of 100 epochs (with 1000 batches, i.e., 2000 patches, per epoch). The learning rate was $10^{-4}$ and kept unchanged throughout training. At the end of each epoch, the trained model was evaluated on validation data and the one with the lowest PEV cosine similarity error (ECSE) among all epochs was selected as the final model. Phase measurements from 6 head orientations were used to train the network, even though during testing, the network can be applied for an arbitrary number of phase inputs. The head orientations were sampled uniformly from within $25^\circ$ with respect to the main magnetic field for training. For each subject, a total of 20 head orientations were sampled, and all possible combinations of 6 orientations were used as training data. In each fold of cross validation, 5 subjects were used for training, 1 for validation and 2 for testing.  Training data include two levels of spatial resolution: low-resolution ($1.5$ mm isotropic) and high-resolution ($0.98\times 0.98 \times 1$mm), allowing for mixed multi-resolution training, with a two-stream batch sampler to ensure each batch contains samples of the same resolution for faster training. The training data of each fold always included two low-resolution subjects and three high-resolution subjects.

The number of iterations for learned proximal updates, $K$, is $4$ in eq. 12. 

In the STI phantom generation, $\Delta$ in eq. 15 was uniformly sampled between $[0,0.002]$ to represent the difference between the two smaller eigenvalues of the susceptibility tensor. $\gamma$ in $a_S = \gamma a_D$ is chosen as $1/15$ to map DTI FA to the range of STI anisotropy.

\section{Complementary Figures}

\begin{figure}[!h]
    \centering
    \includegraphics[trim=0 190 0 0,clip,width=\linewidth]{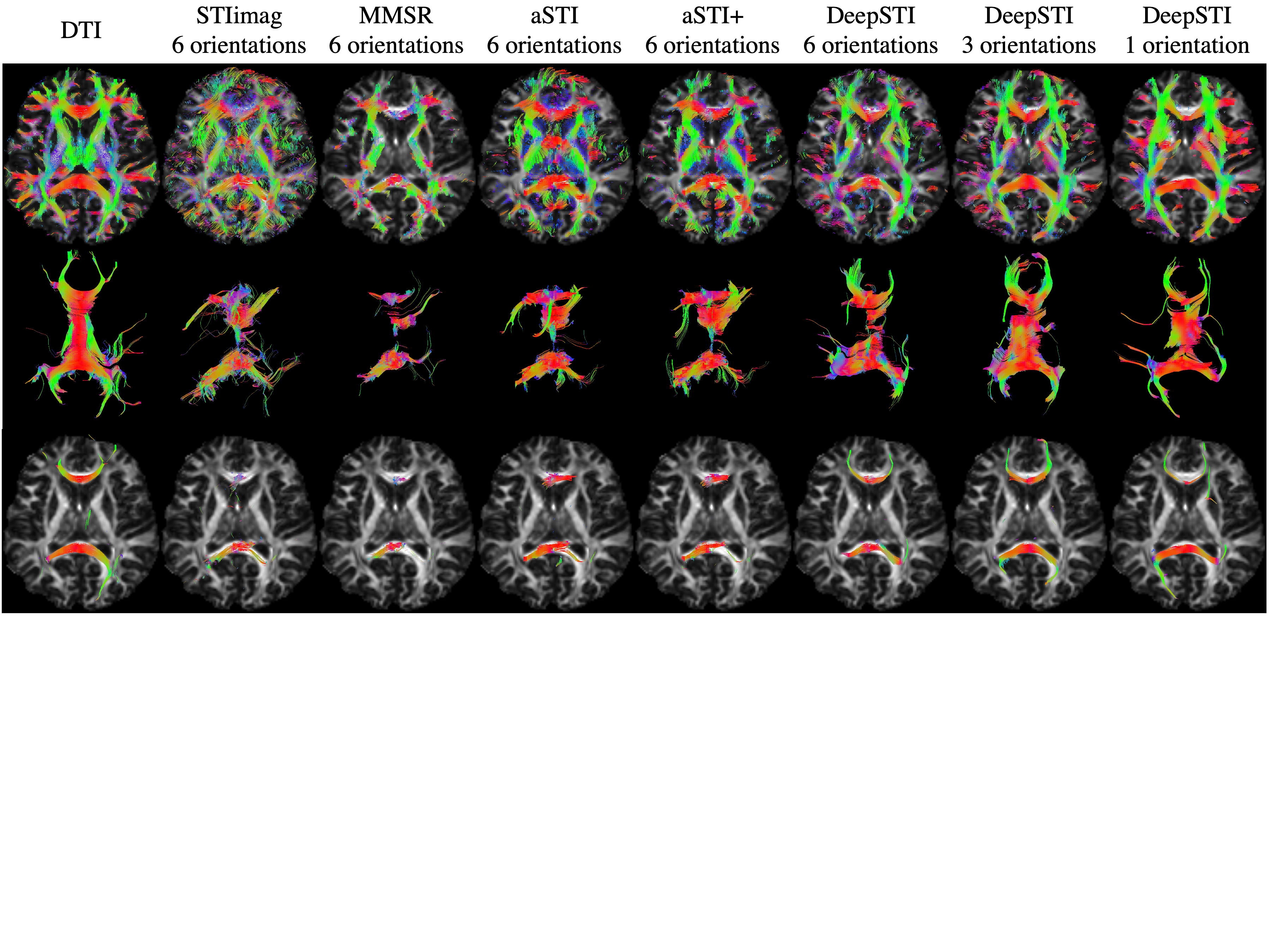}
    \caption{Fiber tractography results of STI from real measurements of a human subject at 3T with 1.5 mm isotropic resolution. Top row: 2D view of whole brain tracking result overlaid on DTI FA. Middle row: 3D volume rendering of neural fibers passing corpus callosum. Bottom row: 2D visualization of neural fibers passing corpus callosum. Columns from left to right: DTI, STIimag result from 6 orientations, MMSR result from 6 orientations, aSTI result from 6 orientations, aSTI+ result from 6 orientations, DeepSTI results from 6, 3 and 1 orientations.}
    \label{fig:res-track-real-3T}
\end{figure}

\section{Detailed Head Orientations}
\begin{table*}[!h]
    \centering
    \begin{tabularx}{\textwidth}{cX}
        \hline\hline
         Number of orientations & Directions of main magnetic field in the subject frame of reference at different orientations \\
         \hline
         6 & (-0.0010, -0.0250, 0.9997), \newline (0.1196, 0.2541, 0.9597), \newline (0.0854, -0.2788, 0.9565), \newline (0.0090, 0.4195, 0.9077), \newline (0.3411, 0.1648, 0.9254), \newline (-0.2203, -0.0452, 0.9744) \\
         \hline
         5 & (-0.0010, -0.0250, 0.9997), \newline (0.1196, 0.2541, 0.9597), \newline (0.0854, -0.2788, 0.9565), \newline (0.0090, 0.4195, 0.9077), \newline (0.3411, 0.1648, 0.9254) \\
         \hline
         4 & (-0.0010, -0.0250, 0.9997), \newline (0.1196, 0.2541, 0.9597), \newline (0.0854, -0.2788, 0.9565), \newline (0.0090, 0.4195, 0.9077) \\
         \hline
         3 & (-0.0010, -0.0250, 0.9997), \newline (0.1196, 0.2541, 0.9597), \newline (0.0854, -0.2788, 0.9565) \\
         \hline
         2 & (-0.0010, -0.0250, 0.9997), \newline (0.1196, 0.2541, 0.9597) \\
         \hline
         1 & (-0.0010, -0.0250, 0.9997) \\
        \hline\hline
    \end{tabularx}
    \caption{Detailed head orientations for \cref{fig:res-pev-real}.}
    \label{tab:head-orientations}
\end{table*}

\end{document}